\documentclass[times,11pt]{elsarticle}
\usepackage{subfigure}
\usepackage{fullpage}
\usepackage{amsmath,amssymb,amsthm,subfig,bm,units,tikz,booktabs}
\usetikzlibrary{arrows}
\usepackage[mathscr]{euscript}
\usepackage{mdframed,booktabs,rotating}
\usepackage[suffix=]{epstopdf}
\usepackage[title,toc,titletoc,page]{appendix}
\graphicspath{{fig/}}

\theoremstyle{definition}

\newtheorem{remark}{Remark}

\DeclareMathOperator*{\argmin}{arg\,min}  
\usepackage{algorithm, algorithmic}

\title{{\bf Physics-informed Data-driven Cavitation Model for a Specific Mie–Grüneisen Equation of State}}

\author[sjtu]{Minsheng Huang}
\ead{mingo.stemon@sjtu.edu.cn}
\author[nint]{Chengbao Yao}
\ead{yaocheng@pku.edu.cn}
\author[nint]{Pan Wang}
\author[sjtu]{Lidong Cheng \corref{cor}}
\ead{critters@sjtu.edu.cn}
\author[sjtu,ins]{Wenjun Ying \corref{cor}}
\ead{wying@sjtu.edu.cn}

\cortext[cor]{Corresponding author}
\address[sjtu]{School of Mathematical Sciences, Shanghai Jiao Tong University, Shanghai, P.R.China}
\address[nint]{Northwest Institute of Nuclear Technology, Xi'an, China}
\address[ins]{MOE-LSC and Institute of Natural Sciences,
Shanghai Jiao Tong University, Shanghai, P.R.China}

\begin{document}
\begin{abstract}
We present a novel one-fluid cavitation model of a specific Mie-Grüneisen equation of state(EOS), named polynomial EOS, based on an artificial neural network. Not only the physics-informed equation but also the experimental data are embedded into the proposed model by an optimization problem. The physics-informed data-driven model provides the concerned pressure within the cavitation region, where the density tends to zero when the pressure falls below the saturated pressure. The present model is then applied to computing the challenging compressible multi-phase flow simulation, such as nuclear and underwater explosions. Numerical simulations show that our model in application agrees well with the corresponding experimental data, ranging from one dimension to three dimensions with the $h-$adaptive mesh refinement algorithm and load balance techniques in the structured and unstructured grid. 
\end{abstract}

\begin{keyword}
Multi-phase flow, One-fluid cavitation model, Artificial neural network, Deep learning, Mie-Gr{\"u}neisen EOS, Underwater explosion.
\end{keyword}

\maketitle

\section{Introduction}
Cavitation in fluid flow occurs when liquid pressure drops to the vapor pressure limit, forming a cavitation zone with a dynamic interface (vapor bubbles), cavitation development, and violent collapse \cite{Arndt1981, Rachid2003}. Unsteady cavitation can be encountered in various practical situations, such as lithotripsy, hydraulic machines, and underwater explosions, where numerous scholars have been extensively researched in recent years, including experimental and numerical methods \cite{liu2003underwater, chen2022underwater, Rajendran2001, Jafarian2017cavitation, Marongiu2010sph, kondama1998}. 

Various numerical models and methods have been proposed to simulate cavitation. These methods can be classified into interface-tracking methods and interface-capturing methods \cite{stavropoulos2021reviewcaviattion}. In the interface-tracking method, a distinct interface is assumed to exist between the liquid and the vapor. This method uses a set of markers or line segments to split cells or deform the mesh in a way that aligns with the interface. The position of the interface is determined by an interactive procedure \cite{Chen1994numeircal, Deshpande1994cavity, Deshpande1997numeircal, liu2004isentropic, stavropoulos2021reviewcaviattion}. On the other hand, the interface-capturing method does not attempt to track the interface between the liquid and vapor. Instead, it treats the flow as a mixture of two phases with a continuously varying mixture density between liquid and vapor \cite{Saurel2001Multiphase, Saurel2008Modelling, Saurel2009Simple, Saurel2018Diffuse, ventikos2000cavitation, Ahuja2001cavitation, liu2004isentropic, schmidt1999fully, xie2005numerical}.

The two-phase method is an efficient way to simulate compressible cavitation flow as it allows for more physical procedures to be considered. There are two types of interface-capturing methods: two-fluid and one-fluid models. In the two-fluid model, each fluid exists in every grid cell and follows its governing equations. This method is widely used in simulating compressible multi-phase and multi-medium fluids, where the mass, momentum, and energy exchange are incorporated as transfer terms. As a result, two-fluid models are excellent in accurately computing mass and energy exchange, heat transfer, and surface tension effects at the cavitation interface, providing significant insights into fluid dynamics \cite{Saurel2018Diffuse, Zein2010Modeling, Saurel2009Simple}. Based on different assumptions, there are several models such as the seven-equation model \cite{Saurel1999multiphase, Baer1986model}, six-equation model \cite{Saurel2009Simple}, five-equation model \cite{allaire2002five, Kapila2001model}, etc. Although the two-fluid model can easily account for additional physical effects, parameters such as exchange rates and viscous friction between different fluids should be determined experimentally in advance \cite{Ahuja2001cavitation, Kubota1992cavitation, Kunz2000cavitation, Senocak2002pressure}. On the other hand, the one-fluid model treats different fluids as a homogeneous mixture and uses only one set of governing equations to describe fluid motion. However, the challenge with the one-fluid model is to find an appropriate EOS for the mixture to describe the phase change from liquid to vapor \cite{schmidt1999fully, ventikos2000cavitation}. In this way, it is assumed that the fluid mixture is always barotropic, making it possible to simulate the formation, development, and collapse of cavitation. This approach also costs less than the two-fluid model. However, the one-fluid model cannot resolve the detailed physics and phase transitions as accurately as the two-fluid model. In some practical situations, such as underwater explosions, the primary concerns are the cavitation dimensions and pressure surges, where the surface tension, thermal conductivity, fluid viscosity, and turbulence effects are neglected.  

The interest of this paper is on unsteady cavitation, which involves the generation, development, and collapse of dynamic interfaces. This phenomenon is commonly observed in underwater explosions, where the liquid and mixture are assumed to be compressible under high-pressure situations. The one-fluid model governed by inviscid Euler equations is used to model such cavitation flows, as cavitation dimension and pressure surges are of primary concern. In recent decades, some well-known and classic one-fluid models have been proposed to simulate the cavitation of compressible flows. The cut-off model, as proposed in \cite{Aanhold1998Underwater}, resets the pressure to a given value (usually saturation vapor pressure) whenever it is detected to be below the critical level $p_{sat}$, widely used in underwater simulations. In this model, the conservation law is violated, and the nature of the hyperbolic equation becomes nonphysical due to the modified pressure and resulting zero sound speed. 
Tang and Huang studied the 1D inviscid cavitating flow based on the vacuum model \cite{Tang1996}. 
The vacuum model treats the cavitation zone of zero mass whose assumption is physically reasonable \cite{liu2004isentropic}.
However, the vacuum model requires adding or deleting the cavitation interface artificially, which is inefficient and complex, especially in high-dimensional cases. 
Schmidt's model \cite{schmidt1999fully} was developed for high-pressure and high-density flow in a small nozzle. 
The EOS in the mixture region was defined by integrating the Wood's sound speed under the assumption that the mixture fluid is barotropic and isentropic. 
However, Schmidt's model was applied in the case in which the large vapor-to-liquid density ratio ($\rho_{g}/\rho_{l})$ is smaller than $ 10^{-5}$. 
Otherwise, the saturation pressure $p_{sat}^{\prime}$ obtained by mixture EOS is negative. Later, Qin et al. \cite{qin1999direct} incorporated a constant model into Schmidt's model for water hammer modeling. Xie et al. \cite{xie2005numerical, xie2006cavitation_a} modified Schmidt's model by replacing $p_{sat}^{\prime}$ with physical saturation pressure $p_{sat}$. They also set pressure limit $p_{\epsilon} = 10^{-9}$ to ensure pressure positivity, making the model suitable for extensive physical applications.
Liu et al. \cite{liu2004isentropic, xie2006cavitation_b, xie2007cavitation} proposed an isentropic model, where the mixture region was assumed to be homogeneous, compressible, and isentropic. The proposed EOS was based on an iterative method to calculate the desired pressure. 
Causon's model \cite{causon2013finite} was obtained by using Oldenbourg polynomials to consider the influence of caloric and thermal behavior in the cavitation region of the compressible fluid. 
Dumbser et al. \cite{Dumbser2013Efficient} projected the thermodynamic table data onto the space of the polynomial of degree $q$ to interpolate the analytical EOS of n-heptane and water. 
Jafarian et al. \cite{jafarian2017exact} presented the model of the cavitation region during the isothermal process, where the fluid phase changes were divided into the wave pattern of a Riemann solver. 
Fu \cite{fu2021} developed a cavitation model based on the general Mie-Gr\"uneisen equation, in which an appropriate Riemann problem and corresponding Riemann solver are developed and applied to underwater explosion simulation. 

The cavitation models discussed above either rely on reasonable physical assumptions or only on the experimental data. However, the physical equation deduced from the physical assumption is not usually consistent with experimental data due to model errors, numerical errors, and measurement errors. The purely data-driven model ignoring the physical assumption is restricted by the range of data and lack of generalization abilities. 
In recent years, deep learning has become popular and has been applied extensively in simulating physical systems, especially as a fitting function or surrogate model. The deep learning method directly transfers the ordinary problem(no matter solving physical equations or fitting data) as an optimization problem by neural networks.
For example, PINN (physics-informed neural network), as explored in several studies \cite{Raissi2019pinn, lulu2021deepxde, pang2019FPINN, Jagtap2020XPINN, ma2023apnn}, is particularly adopted in modeling complex systems by integrating physical principles and initial/boundary conditions. In addition to physics-informed models, data-driven models based on deep learning are also widely used. For example, Tompson et al. \cite{tompson2022accelerating} developed a data-driven method to accelerate solving incompressible Euler equations. In contrast, Liu et al. \cite{liu2020weno} created an artificial neural network to refine the non-linear weights in the WENO-JS scheme. Feng \cite{feng2019shockindicator, feng2021shockindicator} discovered shock indicators in both Cartesian and unstructured grids, revealing the connection between characteristic values and shock generation in an artificial neural network. 

This paper proposes a one-fluid cavitation model of liquid for a typical Mie-Gr\"uneisen equation, named polynomial EOS, by a physics-informed data-driven method based on an artificial neural network. Unlike the abovementioned models, our model relies neither purely on analytical nor data-driven agent models. Based on the framework in PINN, the loss is interpreted as training loss, including initial and boundary loss and residual loss coming from governed equations. Under the assumption that the liquid-vapor mixture state is isentropic and homogeneous, our model is controlled by an ordinary differential equation(ODE) in the cavitation region. Furthermore, experimental data from the Seasame library is incorporated into the model to improve its generality and interoperability. 
Taking the data into account by adding a regularization term to the ODE makes it impossible to solve analytically.
The approximate solution is obtained by an optimization problem consisting of ODE and data loss. The resulting model developed here enhances the capacity of cavitation description and allows us to simulate problems with highly nonlinear fluid involving complex behaviors, such as underwater explosions near a free surface and a solid object.

The remainder of this paper is organized as follows. Mathematical models are introduced in section \ref{sec:mathmodel}. In section \ref{sec:unsteadymodel}, after a brief review of the previous one-fluid models, the details of our new model are presented. Numerical method and algorithm are presented in section \ref{sec:numerical:methods}. In section \ref{sec:num}, several numerical results and discussions are presented compared to other methods, followed by a concluding remark in section \ref{sec:conclusion}.


\section{Mathmetical model}\label{sec:mathmodel}
\subsection{Governing equations}
In this work, gas and liquid are assumed to be compressible due to high-pressure situations. We focus on Euler equations of inviscid single and two-phase flows. Here, ``single'' means the one-fluid model governed by Euler equations. For the sake of simplicity, the mathematical structure can be described according to the following generic representation \cite{schmidmayer2023UEq}: 
\begin{equation}
    \frac{\partial \boldsymbol{Q}}{\partial t}+\nabla \cdot \boldsymbol{F}(\boldsymbol{Q}) + \bm{h}(\bm{Q}) \cdot \nabla \bm{u}=\bm{S}(\bm{Q}), \label{model:hyper}\end{equation}
where $\bm{Q}$ is the vector of evolution variables(conservative or non-conservative), $\bm{u}$ is the mixture velocity field, $\bm{F}$ is a flux function, $\bm{h}$ is the non-conservative quantities, and $\bm{S}$ is the source term. In this paper, we consider the Euler equations and two-phase inviscid compressible flows, where eq. \eqref{model:hyper} is concretized as follows:
\begin{itemize}
    \item The Euler equations \\ 
    The one-fluid model regards the liquid and vapor as a single material governed by a specific EOS. The inviscid compressible flows are modeled by Euler equations, which consist of equations for the conservation of mixture density, momentum, and energy:
    \begin{equation}
        \bm{Q} = \begin{pmatrix}
            \rho\\
            \rho\bm{u}\\
            E
        \end{pmatrix}, \quad \bm{F}(\bm{Q}) = \begin{pmatrix}
            \rho \bm{u}\\
            \rho \bm{u} \otimes \bm{u} + p \bm{I}\\
            (E+p)\bm{u}
        \end{pmatrix}, \quad \bm{h}(\bm{Q}) = \begin{pmatrix}
            0\\
            \bm{0}\\
            0
        \end{pmatrix}, \quad \bm{S}(\bm{Q}) = \begin{pmatrix}
            0\\
            \bm{0}\\
            0
        \end{pmatrix}. 
        \label{model:euler}
    \end{equation}
    Here $\rho, \bm{u}, p$ is density, velocity filed, and pressure filed, respectively, and $E$ stands for the flow total energy per unit volume such that
    \begin{equation*}
        E = \rho e + \frac{1}{2} \rho ||\bm{u}||^{2},
    \end{equation*}
    in which $e$ is the specific internal energy. 

    \item The five-equation model for two-phase inviscid compressible flows\\
    Under the assumption that the pressure and velocity at the material interface reach equilibrium, the five-equation model consists of two-phase mass, one mixture momentum equation, one mixture total energy equation, and an equation for the transport of the volume fraction, which is given as follows
\begin{equation}
    \bm{Q} = \begin{pmatrix}
    \rho_{1}\alpha_{1}\\
    \rho_{2}\alpha_{2}\\
    \rho \bm{u}\\
    E\\
    \alpha_{1}\\
    \end{pmatrix}, \quad \bm{F(\bm{Q})} = \begin{pmatrix}
    \rho_{1}\alpha_{1}\bm{u}\\
    \rho_{2}\alpha_{2}\bm{u}\\
    \rho \bm{u} \otimes \bm{u} + p\bm{I}\\
    (E + p)\bm{u}\\
    \alpha_{1}\bm{u} 
    \end{pmatrix}, \quad \bm{h}(\bm{Q}) = \begin{pmatrix}
        0\\
        0\\
        \bm{0}\\
        0\\
        -\alpha_{1}
    \end{pmatrix},\quad \bm{S}(\bm{Q}) = \begin{pmatrix}
    0\\
    0\\
    \bm{0}\\
    0\\
    0\\
    \end{pmatrix},
    \label{model:five_equation}
\end{equation}
where $\rho_{k}, \alpha_{k}, k \in \left\{1, 2\right\}$ are the density and volume fraction of $k^{th}$ fluid respectively. The volume fraction $\alpha_{k}$ is always non-negative and takes a value between 0 and 1. Different from \eqref{model:euler}, here $\bm{u}$ is the velocity field, $p$ is mixture pressure, and $E$ is total mixture energy per unit volume. One can extend the two-phase model to the arbitrary number of phases by adding a continuity equation and a new advection equation of volume fraction for the additional phase \ {deng2018bvd}.
\end{itemize}

\subsection{Equation of the state}
To close the system, each phase of fluids is assumed to be governed by the general Mie-Geüneisen EOS \cite{shyue2001}:
\begin{equation}
    p = \Gamma \rho e + h(\rho),
    \label{model:eos:basic}
\end{equation}
where $\Gamma, e, h(\rho)$is the Grüneisen parameters, internal energy, and density function, respectively. In this paper, we consider the four specific types of Mie-Grünensen EOS: ideal gas EOS, Tait's EOS, JWL EOS, and polynomial EOS \cite{Jha2014polynomial,fu2021}, which follows as:
\begin{itemize}
    \item \textbf{Ideal gas EOS}\\
    In general, most gases are assumed as ideal gases and \eqref{model:eos:basic} can be reformulated by:
    \begin{equation}
        p = (\gamma - 1) \rho e,
        \label{model:eos:ideal}
    \end{equation}
    where $\Gamma = \gamma - 1$ and $h(\rho) = 0$,$\gamma$ is the ratio of the specific heats and is set to 1.4 unless stated otherwise. The sound speed associated with the ideal gas \eqref{model:eos:ideal} is expressed by
    \begin{equation}
        a_{ideal} = \sqrt{\frac{\gamma p}{\rho}}.
        \label{model:eos:ideal:sound}
    \end{equation}
    \item \textbf{Tait's EOS}\\
    Tait's EOS is widely used for barotropic flow and commonly for describing the thermodynamic behavior of the water in relatively not high pressure \cite{liu2004isentropic}(20000 atomoshpere \cite{glass1994nonstationay}), which gives rise to the relation between pressure and density as
    \begin{equation}
        p = B\left(\frac{\rho_{\ell}}{\rho_{\ell_{0}}}\right)^{N} - B + A,\label{model:tait}
    \end{equation} 
    where $\Gamma = 0$. Since the first term of \eqref{model:eos:basic} disappears, only the density function works. The values of parameters are $N = 7.15, B = 3.31 \times 10^{8}$ Pa, $A = 1.0 \times 10^{5}$ Pa, respectively, and $\rho_{0} = 1000 $ kg/$\text{m}^{3}$ for water. $\bar{p} = p + \bar{B}, \bar{p}_{0} = p_{0} + \bar{B}, \bar{B} = B - A, p_{0} = A$. 
    The sound speed of Tait's EOS is as follows: 
    \begin{equation}
        a_{Tait's} = \sqrt{N \frac{\bar{p}}{\rho_{l}}}.
        \label{model:eos:tait}
    \end{equation}
    \item \textbf{JWL EOS}\\
    The Jones-Wilkins-Lee(JWL) EOS \cite{Lee1968jwl} is widely used to characterize the hydro simulations of thermodynamics of high explosives such as TNT products \cite{Baudin2010JWL}, which is given as:
    \begin{equation}
        p = A_{1} \left(1 - \frac{\omega \rho}{R_{1}\rho_{0}}\right)exp\left(-\frac{R_{1}\rho_{0}}{\rho}\right)  + A_{2}\left(1 - \frac{\omega \rho}{R_{2}\rho_{0}} \right)exp\left(-\frac{R_{2}\rho_{0}}{\rho}\right)+ \Gamma \rho e, 
        \label{model:eos:jwl}
    \end{equation}
    where $A_{1}, A_{2}, R_{1}, R_{2}, \rho_{0}, \Gamma$ are positive parameters determined empirically. In the numerical simulation of this work in section \ref{sec:num}, we describe the TNT products referred by \cite{Smith1999AUSM} as $A_{1} = 3.712\times10^{11} \text{ Pa}, A_{2} = 3.230\times10^{9} \text{ Pa}, R_{1} = 4.15, R_{2} = 0.95, \rho_{0} = 1630 \text{ kg} / \text{m}^{3}, \Gamma = 0.30$ respectively. The sound speed of \eqref{model:eos:jwl} is given by
    \begin{equation}
        a_{JWL} = \sqrt{\frac{(\Gamma + 1)p}{\rho} - A_{1}\frac{\rho + R_{1}\rho_{0} + \Gamma \rho}{\rho^{2}}exp\left(-\frac{R_{1}\rho_{0}}{\rho}\right) - A_{2} \frac{\rho+R_{2}\rho_{0}+\Gamma\rho}{\rho^{2}}exp\left(-\frac{R_{2}\rho_{0}}{\rho}\right)}
    \end{equation}
    
    \item \textbf{Polynomial EOS}\\
    The Polynomial EOS is a specific form of the Mie-Grüniesen EOS, which is widely used to model water in extreme situations and has different states in compression and tension:
    \begin{equation}
        p= \begin{cases}A_1 \mu+A_2 \mu^2+A_3 \mu^3+\left(B_0+B_1 \mu\right) \rho_0 e, & \mu>0, \\ T_1 \mu+T_2 \mu^2+B_0 \rho_0 e, & \mu \leq 0,
        \end{cases}
        \label{model:eos:polynomial}
    \end{equation} 
    where $\mu = \frac{\rho}{\rho_{0}} - 1$, $\rho_{0}$ is the reference density, and other parameters are positive constants. The liquid is in the compression stage when 
    $\rho \geq \rho_{0}$ whlie in tension when $\rho \leq \rho_{0}$. In present study, the parameters are taken for water \cite{Jha2014polynomial, fu2021}:$\rho_{0} = 1000$ kg/$\text{m}^{3}$, $A_{1} = 2.20 \times 10^{9}$ Pa, $A_{2} = 9.54 \times 10^{9}$ Pa, $A_{3} = 1.45 \times 10^{10}$ Pa, $B_{0} = B_{1} = 0.28, T_{1} = 2.20 \times 10^{9}$ Pa and $T_{2} = 0.$ In this case, eq. $\eqref{model:eos:polynomial}$ can be simplified as 
    \begin{equation}
        p= \begin{cases}A_1 \mu+A_2 \mu^2+A_3 \mu^3+B_0 \rho e, & \mu>0, \\ T_1 \mu+B_0 \rho_0 e, & \mu \leq 0.\end{cases}\label{model:eos:polynomial:water}
    \end{equation} 
    The sound speed of the water polynomial EOS \eqref{model:eos:polynomial:water} is given by:
    \begin{equation}
        a_{poly} = \begin{cases} 
        \sqrt{\frac{(A_{1} + 2A_{2}\mu + 3A_{3}\mu^{2})}{\rho_{0}} + B_{0}(e + \frac{p}{\rho})}, & \mu > 0, \\
        \sqrt{\frac{p + T_{1} + p B_{0}}{\rho}}, & \mu \leq 0.
        \end{cases} \label{model:eos:ploysoundspeed}
    \end{equation}   
    
    \item \textbf{Equation of state for five-equation model}\\
    Assumed two-phase flows both have Mie-Grüniesen EOS \eqref{model:eos:basic}, the volume fraction of each phase satisfies the saturation restriction:
    \begin{equation}
        \alpha_{1} + \alpha_{2} = 1.
    \end{equation}
    Conservative constraints define mixture density and mixture internal energy:
    \begin{equation}
    \begin{aligned}
        \rho & = \alpha_{1}\rho_{1} + \alpha_{2}\rho_{2}, \\
        \rho e & = \alpha_{1}\rho_{1}e_{1} + \alpha_{2}\rho_{2}e_{2}.
    \end{aligned}
    \end{equation}
    
    Derived from \cite{abgrall1996pressure, shyue2001}, the resulting mixture EOS reads,
    \begin{equation}
    p(\rho, e, \alpha_{1}, \alpha_{2}) = \left(\sum_{k=1}^{2}\alpha_{k}\rho_{k}e_{k} + \sum_{k=1}^{2}\frac{\alpha_{k}h_{k}(\rho)}{\Gamma_{k}}\right) \bigg{/} \sum_{k=1}^{2}\frac{\alpha_{k}}{\Gamma_{k}}.
    \label{model:mixeos}
    \end{equation}
    
    The corresponding sound speed of $\eqref{model:mixeos}$ is given by Wood's sound speed \cite{Wallis2020}
    \begin{equation}
    \frac{1}{\rho a^{2}_{wood}} = \frac{\alpha_{1}}{\rho_{1}a_{1}^{2}} + \frac{\alpha_{2}}{\rho_{2}a_{2}^{2}}.
    \label{model:wood}
    \end{equation}
    The $\rho, \bm{u}, p$ is the density, velocity, and pressure of the mixture, respectively.
    In the cavitation region, liquid and cavitating flows are assumed to be compressible, where the pressure and density are defined as
    \begin{align}
    p & = p(\rho), \label{model:cav:eos} \\
    \rho &= \alpha \rho_{g} + (1 - \alpha) \rho_{\ell}. \label{model:cav:alpha}
    \end{align}
    Here $\rho$ is the pure averaged mixture density of cavitating flow. $\alpha$ is the void fraction, $\rho_{g}, \rho_{\ell}$ are the densities of vapor and liquid components, respectively. Eq.\eqref{model:cav:eos} is the EOS to describe the cavitating flow for the closure of the system, which is the interest of this paper and will be discussed detailly in section \ref{sec:cavition}.
\end{itemize}


\begin{remark}
    Tait's EOS reveals that the pressure only depends on density, while the others depend not only on density but also on specific internal energy. This barotropic relation excludes the energy equation of \eqref{model:euler}. For more details, readers are recommended to see \cite{liu2004isentropic, xie2006cavitation_a}.
\end{remark}
\section{Unsteady cavitation models}\label{sec:unsteadymodel}
Cavitation occurs when the pressure of the liquid falls below the saturated pressure. Generally, pressure will tend to be negative unless a cavitation model is applied, which results in physical variables losing their positivity and codes broken down. 
Various cavitation models are developed to avoid these nonphysical results, such as the cut-off model, Schmidt's model and its variants, and Liu's isentropic model. In the following subsections, we will 
briefly review these classic and widely used one-fluid models. Then we demonstrate details of our novel physics-informed data-driven cavitation model in the second subsection, which is based on polynomial EOS. 




\subsection{Brief review of classic cavitation model}
\subsubsection{Cut-off model}\label{sec:unsteady:cutoff}
The cut-off model \cite{Aanhold1998Underwater} is one of the most classical cavitation models in simulating underwater explosions. The cut-off model sets the cavitation pressure as a given value, which equals physical saturated vapor pressure when the flow pressure fails below a fixed pressure level. The specific EOS for this model can then be written as 
\begin{equation}
    p = \begin{cases}
    f(\rho), & \rho > \rho_{\text{sat}}, \\
    p_{\text{sat}}, & \rho \leq \rho_{\text{sat}}.
    \end{cases}
\end{equation}
Here, $\rho_{\text{sat}}$ is the liquid density with respect to the critical pressure $p_{\text{sat}}$. $f(\rho)$ is the function of liquid EOS, which only depends on density $\rho$ under the assumption of barotropic flow, such as Tait's EOS in $\eqref{model:tait}$. The cut-off model is straightforward to implement and has been widely used in engineering applications. However, mass loss occurs and results in the violation of conservation law. 

\subsubsection{Schmidt's model and modified Schmidt's model}\label{sec:unsteady:schmidt}
Schmidt's model \cite{schmidt1999fully} is another popular cavitation model to simulate the small nozzles with high-speed and low vapor-liquid density ratio(nearly smaller than $10^{-5}$). Schmidt's model is a one-fluid model obtained assuming that cavitation flow is homogeneous and consists of vapor and liquid. Deduced from Wood's formula in \eqref{model:wood} with the assumption of sound speed of pure phases are constant, Schmid's model is given as:

\begin{equation}
    p=p_{\text {sat }}^{\prime}+p_{\mathrm{g\ell}} \cdot \ln \left[\frac{\rho_{\mathrm{g}} \cdot a_{\mathrm{g}}^2 \cdot\left(\rho_{\ell}+\alpha\left(\rho_{\mathrm{g}}-\rho_{\ell}\right)\right)}{\rho_{\ell} \cdot\left(\rho_{\mathrm{g}} \cdot a_{\mathrm{g}}^2-\alpha\left(\rho_{\mathrm{g}} \cdot a_{\mathrm{g}}^2-\rho_{\ell} \cdot a_{\ell}^2\right)\right)}\right], 
   \quad \rho_{g} \leq \rho \leq \rho_{\ell} ,\label{model:schmidt}
\end{equation}
where $\alpha$ is given by \eqref{model:cav:alpha}, $p_{\mathrm{g\ell}}$ is a constant and not related to $\alpha$, which determined by:
\begin{equation}
    p_{\mathrm{g\ell}}=\frac{\rho_{\mathrm{g}} \cdot a_{\mathrm{g}}^2 \cdot \rho_{\ell} \cdot a_{\ell}^2 \cdot\left(\rho_{\mathrm{g}}-\rho_{\mathrm{\ell}}\right)}{\rho_{\mathrm{g}}^2 \cdot a_{\mathrm{g}}^2-\rho_1^2 \cdot a_{\ell}^2}. \label{model:schmidt:contant}
\end{equation}
To keep positive pressure, the $p_{sat}^{\prime}$ in \eqref{model:schmidt} is not the physical saturated pressure and is set to a larger value.
Later, Xie et.al \cite{xie2006cavitation_a} found that Schmidt's model is unable to simulate the unsteady transient cavitating flow with a large vapor to liquid density ratio since the $p_{sat}^{\prime}$ is set to larger than the physical saturated pressure $p_{sat}$ of incipient cavitation. Assuming that the EOS of the liquid phase is barotropic and employed by Tait's EOS, Xie proposed a novel model for \eqref{model:schmidt} to modify Schmidt's model, which is given as
\begin{equation}
    p = \begin{cases}
        \text{Tait's EOS}, & p \geq p_{sat},\\
        p_{sat} + p_{g\ell} \cdot  \ln \left[\frac{\rho_{\mathrm{g}} \cdot a_{\mathrm{g}}^2 \cdot\left(\rho_{\ell}+\alpha\left(\rho_{\mathrm{g}}-\rho_{\ell}\right)\right)}{\rho_{\ell} \cdot\left(\rho_{\mathrm{g}} \cdot a_{\mathrm{g}}^2-\alpha\left(\rho_{\mathrm{g}} \cdot a_{\mathrm{g}}^2-\rho_{\ell} \cdot a_{\ell}^2\right)\right)}\right], & p_{\epsilon} < p < p_{sat},\\
        p_{\epsilon}, & p \leq p_{\epsilon}.
        \end{cases}
\end{equation}
Here, the $p_{\epsilon}$ is a given small value to prevent cavitation pressure below zero, and $p_{sat}$ is the physical saturated pressure. 


\subsubsection{Liu's isentropic model}\label{sec:unsteady:liu}
Owing to the sound speed formulation in \eqref{model:wood} used for Schmidt's model is mathematically reasonable and verified experimentally, Liu still uses this principle in the cavitation flow. Compared to Schmidt's model, Liu assumes that vapor and liquid are both isentropic and barotropic. The void fraction $\alpha$ can only be a sole function of pressure due to $\eqref{model:cav:alpha}$, which is governed by:
\begin{equation}
    \frac{\mathrm{d} \alpha}{\mathrm{d} p}\left(\rho_{\ell}-\rho_{\mathrm{g}}\right)=\alpha \frac{\mathrm{d} \rho_{\mathrm{g}}}{\mathrm{d} p}+(1-\alpha) \frac{\mathrm{d} \rho_{\ell}}{\mathrm{d} p}-\frac{\mathrm{d} \rho}{\mathrm{d} p}. \label{model:liu}
\end{equation}
The detailed derivation of $\eqref{model:liu}$ can be found in \cite{liu2004isentropic}. After simple manipulation, \eqref{model:eos:tait} and then $\eqref{model:liu}$ can be rewritten as 
\begin{equation}
    \frac{\mathrm{d} \alpha}{\mathrm{d} p}=\alpha(1-\alpha)\left(\frac{1}{\rho_{\ell} a_{\ell}^2}-\frac{1}{\rho_{\mathrm{g}} a_{\mathrm{g}}^2}\right) \label{model:liu:sound1}
\end{equation}
Plugging the sound speed formulation of gas \eqref{model:eos:ideal:sound} and liquid \eqref{model:eos:tait} in \eqref{model:liu:sound1}, then integrating \eqref{model:liu:sound1} from $p_{\text{cav}}$ to $p$, we have
\begin{equation}
    \frac{\alpha}{1-\alpha}=k \frac{\left(\bar{p} / \bar{p}_{\mathrm{cav}}\right)^{1 / N}}{\left(p / p_{\mathrm{cav}}\right)^{1 / \gamma}}
\end{equation}
and 
\begin{equation}
    \rho=\frac{k \rho_{\mathrm{g}}^{\mathrm{cav}}+\rho_{\ell}^{\mathrm{cav}}}{\left(\frac{\bar{p}}{\bar{p}_{\text {cav }}}\right)^{-1 / N}+k\left(\frac{p}{p_{\text {cav }}}\right)^{-1 / \gamma}},
\end{equation}
where $k = \frac{\alpha_{0}}{1-\alpha_{0}}$, $\rho_{g}^{\text{cav}}$ and $\rho_{\ell}^{\text{cav}}$ are the associated gas and liquid density with respect to $p_{\text{cav}}$ respectively. $\alpha_{0}$ is a known void fraction of the mixture density at cavitation pressure $p_{\text{cav}}$. In fact, $p_{\text{cav}}$ and $\alpha_{0}$ is the constant coming from physical experiments. Liu's isentropic model takes advantage of sound speed $\eqref{model:wood}$, which is accurate in the comparison to experimental results \cite{xie2007cavitation,xie2006cavitation_b} and more mathematically consistent with deriving EOS, which is widely used in underwater explosions and other cavitating simulations\cite{xie2007cavitation, xie2006cavitation_b, hu2013cavitation}.

\subsection{Our physical-informed data-driven cavitation model}
\label{sec:cavition}
In this section, we will present our physics-informed data-driven model for the cavitation region. To include the physics into the proposed model, we assume that the mixture in cavitation is isentropic and homogeneous, the same as Liu et al. \cite{liu2004isentropic}. As a result, the deduced equation is the same as \eqref{model:liu:sound1} while is different from the speed of sound due to the different EOS, which will be discussed in the following subsection. Otherwise, the experimental data comes from the Sesame EOS data \cite{sesame, xia1992, xia1993}, which is a tabulated EOS library of more than 150 materials. The experimental data in Sesame checks the consistency of the thermodynamic interpolations in different regions. Furthermore, tabulated equilibrium data \cite{osti_1419738} in logically-cartesian tables is involved to populate the phase-space with other macroscopic properties like pressure and internal energy \cite{2021APSDPPNP1130F}.

\begin{figure}[ht!]
    \centering
    \includegraphics[width=\linewidth]{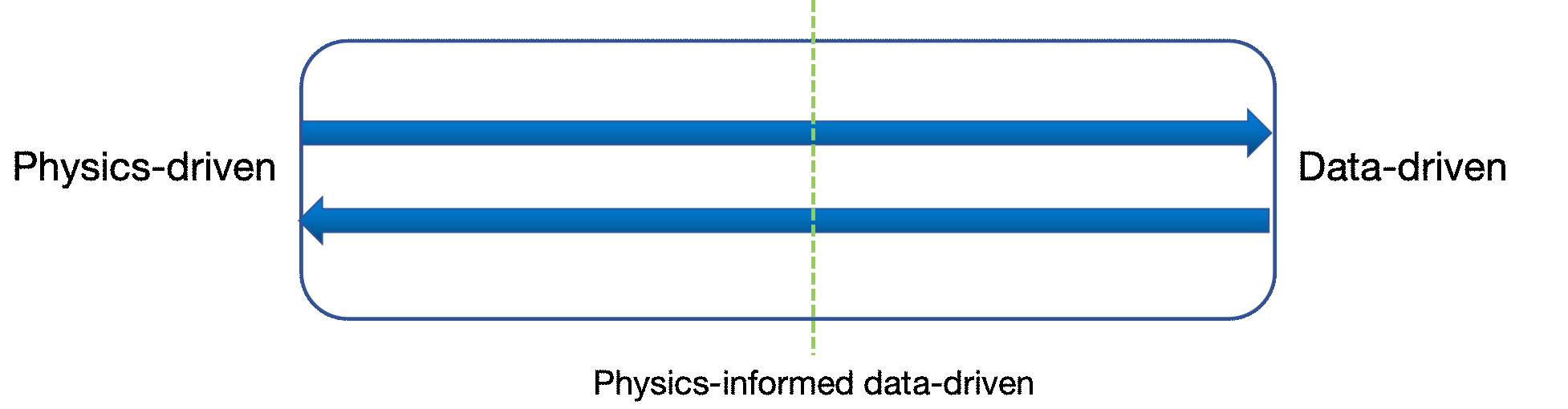}
    \caption{Deep learning in physical simulation.}
    \label{fig:model:physical_inform}
\end{figure}
As shown in Fig. \ref{fig:model:physical_inform}, most physical models are developed by physics-driven or data-driven methods. In the physics-driven method, the deduced model is described by equations, such as ordinary differential equations or partial differential equations. The most important thing is to find the well-posed solution that matches the initial or boundary condition. Unlike the physics-driven method, the data-driven method tends to discover the physical laws behind the data, regardless of the numerical or experimental data. However, the results deduced by analytical or numerical methods of the physical model sometimes cannot satisfy the experimental data due to measurement, numerical, and modeling errors. Relying solely on the data-driven method will cause the physical constraints to be lost or restricted within a limited region, thereby reducing the generalization abilities. To balance the physical equations and experimental data, we regard the experimental data as the supervisor term in the loss function, combining the physical loss and data loss. 

\subsubsection{Modify the physical model with data}
According to the literature \cite{liu2004isentropic}, Wood's sound speed \eqref{model:wood} preserves the 
sound mathematic property and behaves accurately compared to the experimental results \cite{Brennen2013cavitation, Wallis2020} for the homogeneous and thermal equilibrium mixture for neglecting mass or energy transfer on the sound speed and additional physics. 
With the same assumption as Liu's isentropic model, the local sound speed of the mixture follows \eqref{model:wood}, and the void fraction $\alpha$ is controlled by \eqref{model:liu:sound1}. 

The sound speed of polynomial EOS is given as \eqref{model:eos:ploysoundspeed}. It should be pointed out that the cavitation phenomenon happens during the expanding procedure of the liquid. Therefore, only mixture density $\rho \leq \rho_{0}$(or $\mu < 0$) need to be considered. We reformulate \eqref{model:liu:sound1} by substituting 
\begin{equation}
    \alpha = \frac{\rho - \rho_{\ell}}{\rho_{g} - \rho_{\ell}},
\end{equation}
which gives
\begin{equation}
    \left(\frac{d p}{d \rho}\right)^{-1} = \frac{\rho}{\rho_{g} - \rho_{\ell}}\left(\frac{\rho-\rho_{\ell}}{\rho_{g}a_{g}^{2}} + \frac{\rho_{g}-\rho}{\rho_{\ell}a_{\ell}^{2}}\right).
    \label{model:pinn:ode}
\end{equation}
Here, mixture pressure only depends on mixture density in the cavitating flow since $\rho_{g}, \rho_{l}$ is the function of $\rho$. Once given the proper condition $p(\rho_{sat}) = p_{sat}$, the solution \eqref{model:pinn:ode} is well-posed and can be solved analytically or numerically. 

In the Seasame EOS library, the experimental data is arranged in pressure, density, specific internal energy, and temperature. For the sake of simplicity, we consider the barotropic cavitating flow and neglect the influence of temperature. To ensure the compatibility of the physical model with experimental values, we selected the density and pressure of water vapor at a temperature of 290K as our data set, denoted as $\mathcal{D}(\rho, p)$. It is not straightforward to incorporate the experimental data into the \eqref{model:pinn:ode} analytically. Pressure $p$ is not only governed by a physics-informed differential equation \eqref{model:pinn:ode} but also influenced by data set $\mathcal{D}$. The relationship between density and pressure is complex and complicated to be described. However, the neural network provides an alternative approach to achieve this. 

Let the physics-informed neural network be $p(\rho;\bm{\Theta})$, where $\bm{\Theta}$ is the parameter space of the neural networks. We can transform the problem into an optimization problem, which consists of two main parts: one part is the physical equations, and the other is the experimental data. These two components will be described in sequence in the following.

First, we define $f(p, \rho)$ to be 
\begin{equation}
    f(p, \rho) = \left(\frac{d \rho}{d p}\right)^{-1} - \frac{\rho}{\rho_{g} - \rho_{\ell}}\left(\frac{\rho-\rho_{\ell}}{\rho_{g}a_{g}^{2}} + \frac{\rho_{g}-\rho}{\rho_{\ell}a_{\ell}^{2}}\right).
\end{equation}
Then the loss of ODE \eqref{model:pinn:ode} with its initial condition is given as
\begin{equation}
    \text{loss}_{ode}(\bm{\Theta}) = \lambda_{1} \text{loss}(\bm{\Theta})_{interior} + \lambda_{2} \text{loss}(\bm{\Theta})_{initial},
    \label{model:pinn:loss}
\end{equation}
where
\begin{equation}
    \begin{aligned}
         \text{loss}_{interior}(\bm{\Theta}) &= \frac{1}{N_{1}}\sum_{i = 1}^{N_{1}}|f(p_{interior}^{i}, \rho_{interior}^{i};\bm{\Theta})|^{2}, \\
         \text{loss}_{initial}(\bm{\Theta}) &= |p(\rho_{sat};\bm{\Theta}) - p_{sat}|^{2}.
    \end{aligned}
\end{equation}
Here, $\{p_{interior}^{i}, \rho_{interior}^{i}\}_{i=1}^{N_{1}}$ denotes the collection points on $f(p,\rho)$. Consequently, $\text{loss}_{inital}$ corresponds to the loss on the initial data, and $\text{loss}_{interior}$ penalizes the equation not being satisfied on the collection points. $\lambda_{1}, \lambda_{2}$ denotes initial and interior loss weights, respectively. If the initial weights is more constrained, $\lambda_{1}$ should be set larger than $\lambda_{2}$, otherwise $\lambda_{2}$ should be larger one. 

To keep the positivity of the physical quantities, such as density and pressure, and overcome the severe stiffness of \eqref{model:pinn:ode}, we first reformulate the pressure and density as follows
\begin{equation}
    \begin{aligned}
        \hat{\rho} & = \log \rho, \quad \rho = e^{\hat{\rho}}, \\
        \hat{p} &= \log p, \quad p = e^{\hat{p}},
    \end{aligned}
    \label{model:pinn:ode:reform1}
\end{equation}
which gives 
\begin{equation}
    d \hat{\rho} = \frac{1}{\rho} d\rho, \quad d \hat{p} = \frac{1}{p} d p.
    \label{model:pinn:ode:reform2}
\end{equation}
By plugging \eqref{model:pinn:ode:reform2} into \eqref{model:pinn:ode},
\begin{equation}
    \left(\frac{d \hat{p}}{d \hat{\rho}}\right)^{-1} = \frac{e^{\hat{p}}}{\rho_{g} - \rho_{\ell}}\left(\frac{e^{\hat{\rho}}-\rho_{\ell}}{\rho_{g}a_{g}^{2}} + \frac{\rho_{g}-e^{\hat{\rho}}}{\rho_{\ell}a_{\ell}^{2}}\right).
    \label{model:pinn:ode2}
\end{equation}
With the reformulation \eqref{model:pinn:ode:reform1}, the mixture density and pressure firstly are transformed in logarithmic function, and then the neural network trains the $\hat{\rho}, \hat{p}$. The output $\rho, p$ needs inverse transformation by the exponential function, which keeps the positivity of the physical quantities. As density decreases from saturation pressure, pressure decreases rapidly, which leads to the large amount of $\frac{d p}{d \rho}$. To simulate this behavior, we add more random points around $(\rho_{sat}, p_{sat})$. Here, we recommend residual adaptive refinement(RAR) developed by Lu et.al \cite{lulu2021deepxde}, which can efficiently reduce the error of discontinuity.

\begin{figure}[t!]
    \centering
    \includegraphics[width=0.8\textwidth]{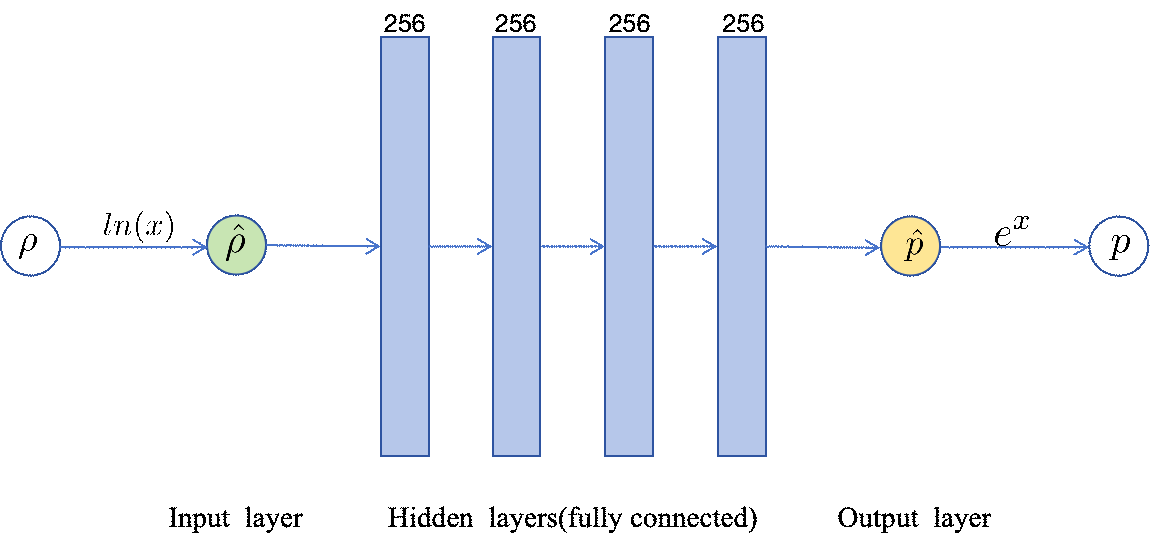}
    \caption{The architecture of the deep learning model used in this work. }
    \label{fig:model:neural_network}
\end{figure}

The loss function in \eqref{model:pinn:loss} consists of physical information, while the experimental data is excluded. The experimental data, denoted as $\mathcal{D}(\rho, p) = \{\rho^{data}, p^{data}\}_{i=1}^{N_{data}}$, consists of density and pressure at temperature 290K, in which the temperature of the physical saturated pressure is most consistent with $(\rho_{sat}, p_{sat})$. Conventionally, the experimental data $\mathcal{D}(\rho, p)$ may not agree with the physical-governed equations due to the disturbances of measuring and unreasonable assumptions on which the equations are based. The physical assumptions should be modified to make the numerical or analytical results more consistent with the experimental data, but this is beyond the scope of the present work. In this paper, we learn physical information from thedata by adding a supervisor term in \eqref{model:pinn:loss}, which follows 
\begin{equation}
    \text{loss}_{total}(\bm{\Theta}) = loss_{ode}(\bm{\Theta}) + \lambda_{3}loss_{data}(\bm{\Theta}),
\end{equation}
where 
\begin{equation}
    \text{loss}_{data}(\bm{\Theta}) = \frac{1}{N_{data}}\sum_{i=1}^{N_{data}}|p(\rho^{i};\bm{\Theta}) - p^{i}|^{2}.
\end{equation}
Here, $\lambda_{3}$ is the weight of data loss, which needs to be selected thoughtfully. If we prefer the experimental data to physical equations, we should set $\lambda_{3}$ much larger than $\lambda_{1}$, otherwise $\lambda_{1}$ should be much larger than $\lambda_{3}$. In comparison to $\lambda_{3}$, $\lambda_{2}$ is only concerned with $\lambda_{1}$, which represents the initial constraints. The final optimization problem can then be written as 
\begin{equation}
    \begin{aligned}
        \bm{\Theta}^{*} &= \argmin_{\bm{\Theta}} \text{loss}_{total}(\bm{\Theta}), \\
        \text{loss}_{total}(\bm{\Theta}) & = \underbrace{\lambda_{1} \text{loss}(\bm{\Theta})_{interior} + \lambda_{2} \text{loss}(\bm{\Theta})_{initial}}_{\text{physics loss}} + \underbrace{\lambda_{3}loss_{data}(\bm{\Theta})}_{\text{data loss}},
    \end{aligned}
\end{equation}
where $\bm{\Theta}^{*}$ denotes a set of (sub)optimal weights obtained from the optimization.

\subsubsection{Architect and training}
The neural network architecture of this work is shown in Fig. \ref{fig:model:neural_network}. We first transform the physical density $\rho$ to $\hat{\rho}$ with the logarithmic function as the neural network input. The network is fully connected and has four hidden layers, each with 256 neurons. The neural network output is not the physical pressure but $\hat{p}$. The real pressure needs to be transformed by the exponential function. The forward(logarithmic) and inverse(exponential) transformations mainly have three advantages: 1) reduce the learning scope since there's no normalization procedure; 2) decrease the stiffness of the \eqref{model:pinn:ode}; 3) ensure the positivity of physical quantities.

\begin{figure}[ht!]
    \centering
    \includegraphics[width=0.5\textwidth]{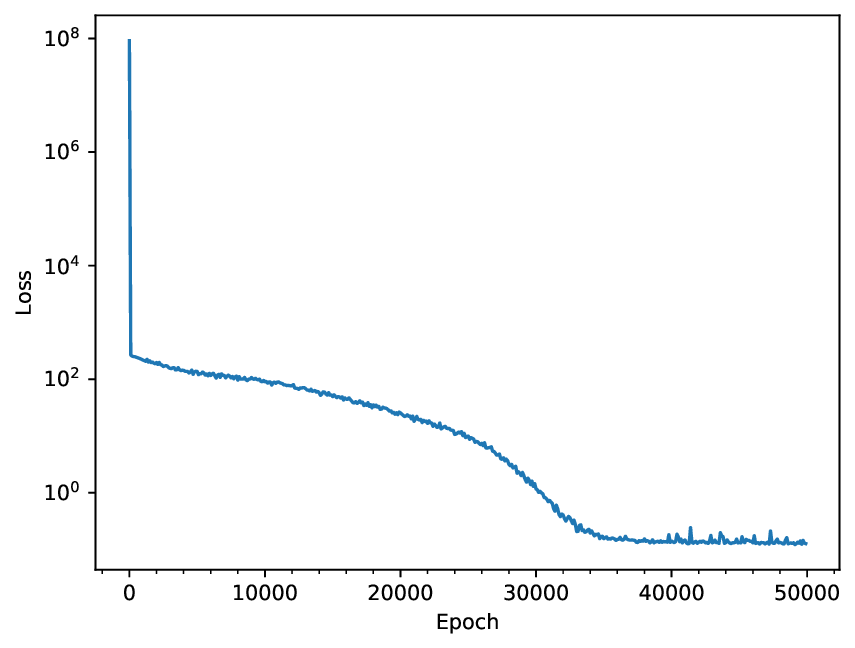}
    \caption{Training loss curve along with epoch.}
    \label{fig:model:training}
\end{figure}

The metrics considered when evaluating the model are the mean square error(MSE). In the training procedure, we prefer physics to experimental data since data is not enough and is affected by the measurement. Therefore, we set $\lambda_{1} = 5.0, \lambda_{2} = 5.0, \lambda_{3} = 0.001$, which shows the best convergence performance. In the initial condition of \eqref{model:pinn:ode}, we set the physical reasonable value according to the experimental data. We use the Adam optimizer to optimize the batch size of 512 and 50000 epochs. The training total loss curve is shown in Fig. \ref{fig:model:training}. 

\subsubsection{The physical-informed data-driven model}
In this subsection, we propose the physical-informed data-driven model for cavitation flow. Let $\mathcal{N}(\rho;\bm{\Theta}^{*})$ be the neural network after training and $\bm{\Theta}^{*}$ be (sub)optimal parameters, the following model and resultant EOS is 
\begin{equation}
    p = \begin{cases}
        A_{1}\mu + A_{2}\mu^{2}+A_{3}\mu^{3}+B_{0}\rho e, & \mu > 0,\\
        T_{1}\mu + B_{0}\rho_{0}e, &\mu \leq 0, p > p_{sat}, \\ 
        \mathcal{N}(\rho;\bm{\Theta}^{*}), &\mu \leq 0, p_{\epsilon} < p < p_{sat}, \\
        p_{\epsilon}, & p < p_{\epsilon},
    \end{cases}
    \label{model:pinn:eos}
\end{equation}
where $p_{\epsilon} = 10^{-9}$ is a given small positive value close to zero, $p_{sat}$ is given physical saturated pressure in experimental data. The cut-off of pressure is to prevent the negative value in the cavitation region. The major difference between our model and previous models can be summarised as follows:
\begin{itemize}
    \item The model is not only dependent on the physical model \eqref{model:pinn:ode}, but also incorporates the data information. In the numerical simulation, the pressure tends to zero when density approaches zero, which is difficult for other methods. 
    \item This model is not fixed but flexible since the different weights$\lambda_{1}, \lambda_{2},$ and $\lambda_{3}$ lead to different models. One adjusts the preference to meet the demand of a practical model, which is more convenient than other methods. 
\end{itemize}

For simplicity, we present the algorithm in a single time step. If we assume that the physical variables at $t = t^{n}$ are known, the algorithm of the physical-informed data-driven model to obtain the solution of liquid and cavitation flow at the next sub-time step is given as follows:

\begin{algorithm}
\caption{Solution procedure at single sub-time step}
\begin{algorithmic}
\REQUIRE $\text{flow variables at $t^{n}$}: \rho^{n}, p^{n}, e^{n}, E^{n}$.      
\ENSURE $\text{flow variables at next sub-time step} \rho^{*}, p^{*}, e^{*}, E^{*}$. 
\STATE 1) Solve the \eqref{model:euler} to obtain the flow variables at the new sub-time step except for the pressure($p$). 
\STATE 2) Evaluate the value of the pressure($p$) and check if pressure fall below the physical saturated pressure($p_{sat}$).
\STATE 3) If $p \geq p_{sat}$, $p^{*}$ is solved by polynomial EOS. Otherwise, \eqref{model:pinn:eos} is employed to evaluate new pressure. 
\end{algorithmic}
\label{algo1}
\end{algorithm}

\begin{remark}
    It is straightforward to extend to another Mie-Grüensen EOS since the main difference is the sound speed in \eqref{model:pinn:ode}. In this work, we trained polynomial EOS, which describes the water more physically when in underwater explosion.
\end{remark}

\section{Numerical methods}\label{sec:numerical:methods}
In this section, we briefly introduce the Godunov-type finite volume method for solving the hyperbolic system \eqref{model:hyper} numerically and then propose the algorithm for simulating with our physics-informed data-driven model. Let a decomposition of the computational domain $\Omega$ into disjoint cells $C_{i}$, which may be associated with a structured or unstructured grid. 
Without loss of generality, we focus on three-dimensional problems in this section, while one and two-dimensional implementations are straightforward. We first integrate the eq.\eqref{model:hyper} over a finite volume cell $C_{i}$,  yielding to 
\begin{equation}
    \frac{d \Bar{\bm{Q}}_{i}}{dt} = \mathcal{L}(\Bar{\bm{Q}}_{i}),
     \label{numerical:semidis}
\end{equation}
where 
\begin{equation}
    \begin{aligned}
      \Bar{\bm{Q}}_{i} &= \frac{1}{V(C_{i})} \int_{C_{i}}\bm{Q}(\bm{x},t^{n}) d\bm{x},\\
      \mathcal{L}(\bm{Q}) &= -\frac{1}{V\left(C_i\right)}\left(\int_{\partial C_{i}} \bm{F}(\bm{Q}) \cdot \bm{n} dSdt + \int_{C_{i}}h(\bm{Q}) \cdot \nabla \bm{u} d \bm{x} dt - \int_{C_i} \bm{S}(\bm{Q}) \bm{d} \bm{x} dt\right),
    \end{aligned}
\end{equation}
with $V(C_{i}), \partial C_{i}$ denoting the volume and the surface of the cell $C_{i}$, respectively. The first term on the right side of the \eqref{numerical:semidis} can be calculated by the summation of all faces of the cells:
\begin{equation}
\int_{\partial C_i} \bm{F}(\bm{Q}) \cdot \bm{n} dSdt = \sum_{j}\int_{\partial C_{i,j}} \bm{F}(\bm{Q}) \cdot \bm{n} \bm{d} Sdt \approx \sum_{j}\bm{F}_{i,j}^{Riemann}(\bm{Q}^{L},\bm{Q}^{R})|C_{i,j}|.
\end{equation}
Here, $\partial C_{i,j}$ denotes the surface between the cell $C_{i}$ and $C_{j}$, $|C_{i,j}|$ is the area of $\partial C_{i,j}$. $\bm{F}_{i,j}^{Riemann}(\bm{Q}^{L}, \bm{Q}^{R})$ is the numerical flux across the edge $\partial C_{i,j}$. $\bm{Q}^{L}, \bm{Q}^{R}$ stand for the left and right-side of values of $\bm{Q}(\bm{x})$ across the $\partial C_{i,j}$. In this paper, the numerical flux $\bm{F}_{i,j}^{Riemann}(\bm{Q}^{L}, \bm{Q}^{R})$ is solved by HLLC Riemann solver \cite{torobook}. The values across the cell boundaries are reconstructed by MUSCL scheme \cite{torobook} with Vanleer slope limiter \cite{vanleer1979slope} unless stated otherwise. 

According to the literature \cite{schmidmayer2023UEq}, let $\bm{Q}_{i}^{n}$  as the solution of $t = t^{n}$, which is calculated or given as the initial condition, resulting to the semi-discretization
\begin{equation}
     \frac{d \Bar{\bm{Q}}_{i}}{dt} = -\frac{1}{V(C_{i})}\left(\sum_{j}\bm{F}_{i,j}^{Riemann}(\bm{Q}^{L}, \bm{Q}^{R})|C_{i,j}|+h(\bm{Q}_{i}^{n})\sum_{j}\bm{u}_{i,j}^{*}|C_{i,j}|-V(C_{i})S(\Bar{\bm{Q}}_{i})\right),
     \label{numerical:semi}
\end{equation}
where $\bm{u}_{i,j}^{*}$ is the flow-velocity vector, which is depreciated to the speed of the contact discontinuity to ensure a correct treatment of the transport equation of $\alpha_{1}$.
Once given the spatial discretization, we employ the two-stage second-order SSP (Strong Stability-Preserving) Runge–Kutta scheme\cite{gottlieb2001ssp} for numerical tests:
\begin{equation}
    \begin{aligned}
    \Bar{\bm{Q}}^{*} & = \Bar{\bm{Q}}^{n} + \Delta t \mathcal{L}(\Bar{\bm{Q}}^{n}), \\
    \Bar{\bm{Q}}^{n+1} & = \frac{1}{2}\Bar{\bm{Q}}^{n} + \frac{1}{2}(\Bar{\bm{Q}}^{*}+\Delta t \mathcal{L}(\Bar{\bm{Q}}^{*})).
    \end{aligned}
    \label{numerical:timeintegral}
\end{equation}
In our model, we need one more step for the cavitation model and calculate the concerted pressure. In summary, the algorithm for simulating compressible flows with our novel model is followed in Algorithm \ref{algo2}.
\begin{algorithm}[htbp]
\caption{Solution procedure at single time step}
\begin{algorithmic}
\REQUIRE $\text{flow variables at $t^{n}$}: \Bar{\bm{U}}^{n}$.      
\ENSURE $\text{flow variables at $t^{n+1}$}: \Bar{\bm{U}}^{n+1}$. 
\STATE 1) Given the cell-average value $\Bar{\bm{U}}^{n}$, then reconstruct the value $\bm{U}^{L}, \bm{U}^{R}$ on cell boundaries via MUSCL scheme. 
\STATE 2) Compute the numerical flux from the reconstructed values across the cell boundary with HLLC Riemann solver.
\STATE 3) Evolve \eqref{numerical:semi} by applying Algorithm.\ref{algo1} to update flow variables in a single sub-time step \eqref{numerical:timeintegral},
\STATE 4) Repeat steps 1) to 3) until the desired sub-time steps are reached, and get $\bm{\Bar{U}}^{n+1}$ at $t^{n+1}$. 
\end{algorithmic}
\label{algo2}
\end{algorithm}

\section{Numerical results}\label{sec:num}
In this section, we present some numerical examples to validate our methods, including one-dimensional cavitation tests and two-dimensional and three-dimensional nuclear/underwater explosion applications. One-dimensional simulations are carried out on a uniform Cartesian grid. In contrast, two-dimensional and three-dimensional simulations are carried out on a mesh composed of quadrilateral and hexahedral cells respectively where the $h$-adaptive mesh refinement(AMR) method and loading balance algorithm are adopted to achieve higher computing efficiency.

\subsection{One-dimensional problems}
In this part, we present some numerical examples of one-dimensional Riemann problems. The computational domain is $[0,1]$ with $400$ uniform cells. The reference solution, if mentioned, is computed on a very fine mesh with $10000$ uniform cells. In our model, we set the $p_{sat} = 2008.445$ Pa, matching with the optimization results of the neural network $\mathcal{N}(\rho;\bm{\Theta}^{*})$. 

\subsubsection{Underwater explosion in one-dimensional spherical free field}
\label{sec::underwater_sphere}
In this part, we present an underwater explosion problem, which is a one-dimensional spherically symmetric problem, and the governing equations are formulated as follows
\begin{equation}
\dfrac{\partial}{\partial t}
\begin{bmatrix}
r^2\rho \\ r^2\rho u \\ r^2 E
\end{bmatrix}
+\dfrac{\partial}{\partial r}
\begin{bmatrix}
r^2\rho u \\ r^2(\rho u^2+p) \\ r^2(E+p)u
\end{bmatrix}
=\begin{bmatrix}
0 \\ 2rp \\ 0
\end{bmatrix}.
\label{eq:spheq}
\end{equation}
The source term in \eqref{eq:spheq} is discretized using an explicit Euler method. 

\begin{figure}[ht!]
\centering
\subfigure[Peak overpressure]
{\includegraphics[width=0.47\textwidth]{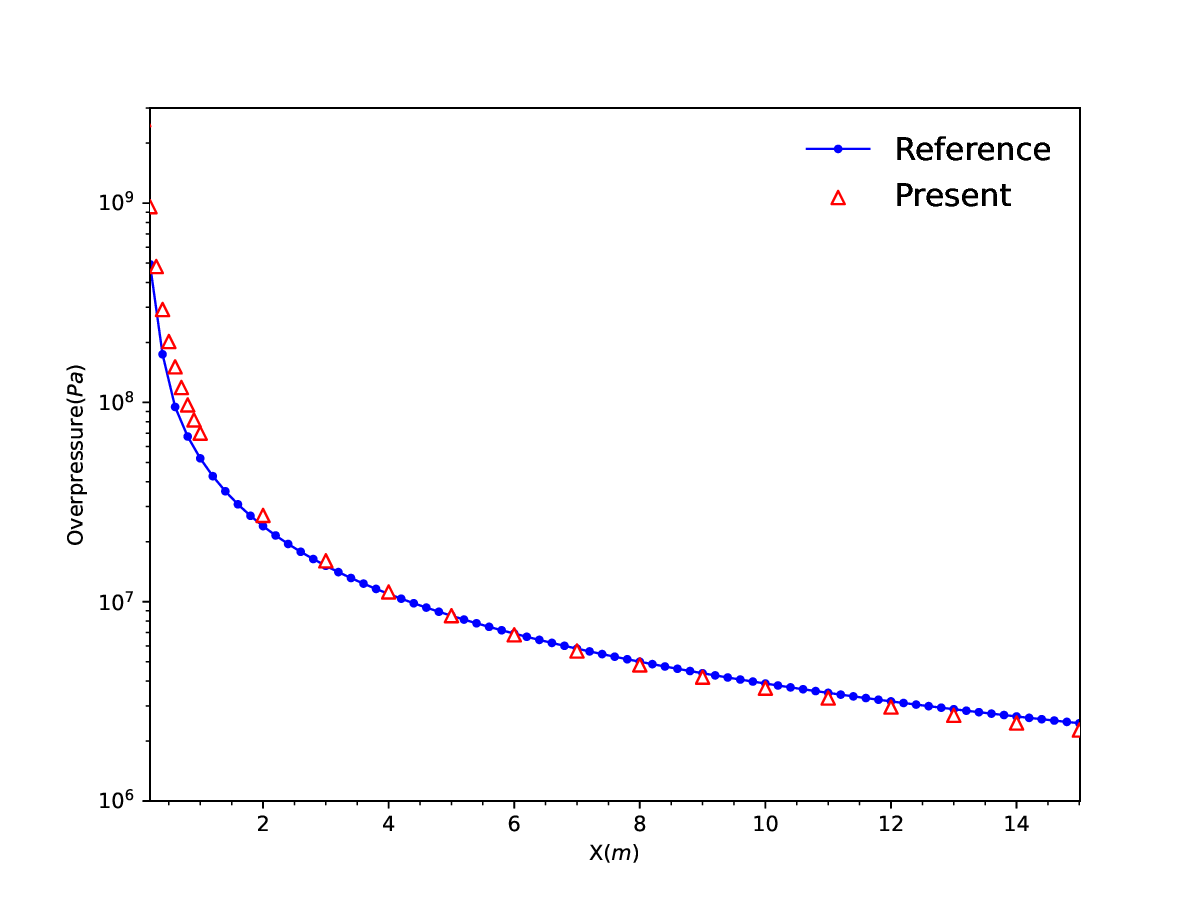}} ~~
\quad
\subfigure[Impulse]
{\includegraphics[width=0.47\textwidth]{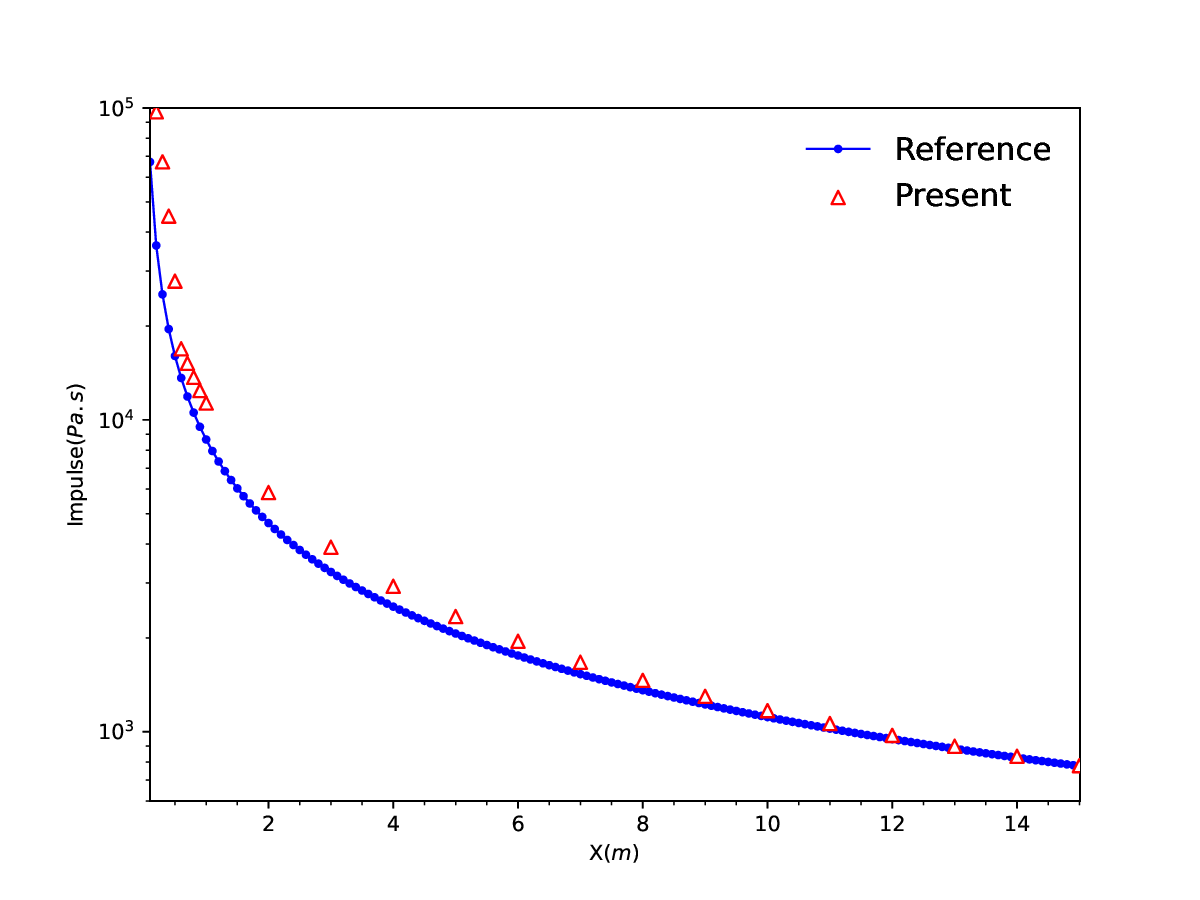}}
\caption{Shock wave parameters for 1.0kg TNT underwater explosion problem.}
\label{rm::udex}
\end{figure}

We use this example to simulate the underwater explosion problem, where a TNT of one kilogram explodes in the water. The high explosives and water are characterized by the JWL EOS and polynomial EOS, respectively. The radii of the computational domain and the initial phase interface are 15m and 0.0527m. The initial internal energy of the high explosives is $4.2\times 10^6$J/kg. Fig.  \ref{rm::udex} shows the computed peak overpressure and impulse at different radii. The results agree well with the empirical data provided in \cite{Cole1948}.

\subsubsection{1D cavitating flow in an open tube in high pressure}\label{numerical:exmaple1}
In this part, we present a Riemann problem of two highly pressurized water flows moving away in opposite directions with the same velocity magnitude from $x = 0.5$. The initial pressure is set to 1000 bar. Two opposite rarefactions are generated and expand from the center if the magnitude of the initial water flows is not high enough, as analyzed in \cite{liu2004isentropic, Tang1996}. To verify the present model, we set different situations and then compared them with the cut-off model, modified Schmidt's model, and Liu's isentropic model. 
\begin{figure}[t!]
    \centering
    \includegraphics[width=0.5\textwidth]{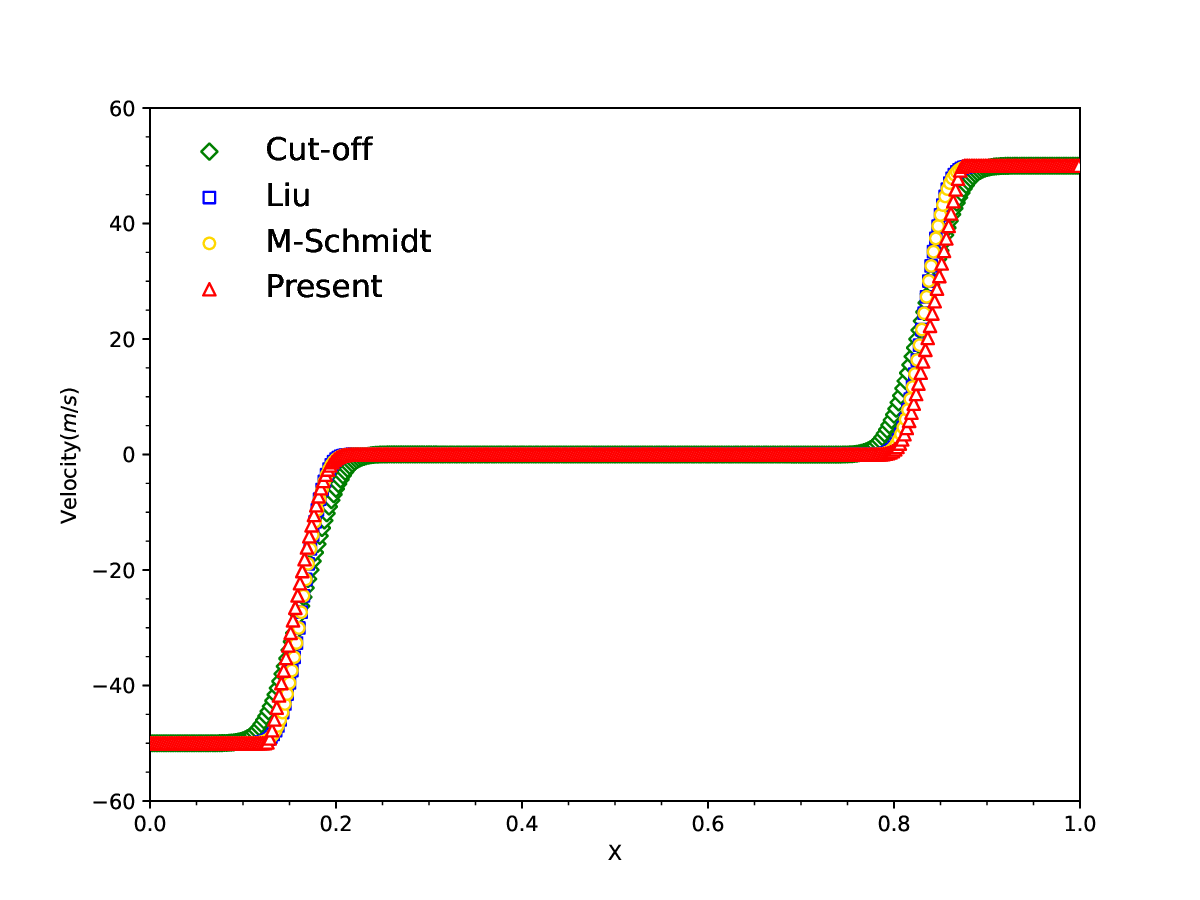}
    \caption{The comparison among the velocity profile of the cutoff model, Liu's isentropic model, modified Schmidt's model, and present model with the initial velocity of $50$m/s at time $t=0.2$ms in the example \ref{numerical:exmaple1}.}
    \label{numerical:example1:1}
\end{figure}

\begin{figure}[ht!]
    \centering
    \subfigure[velocity]{\includegraphics[width=0.49\textwidth]{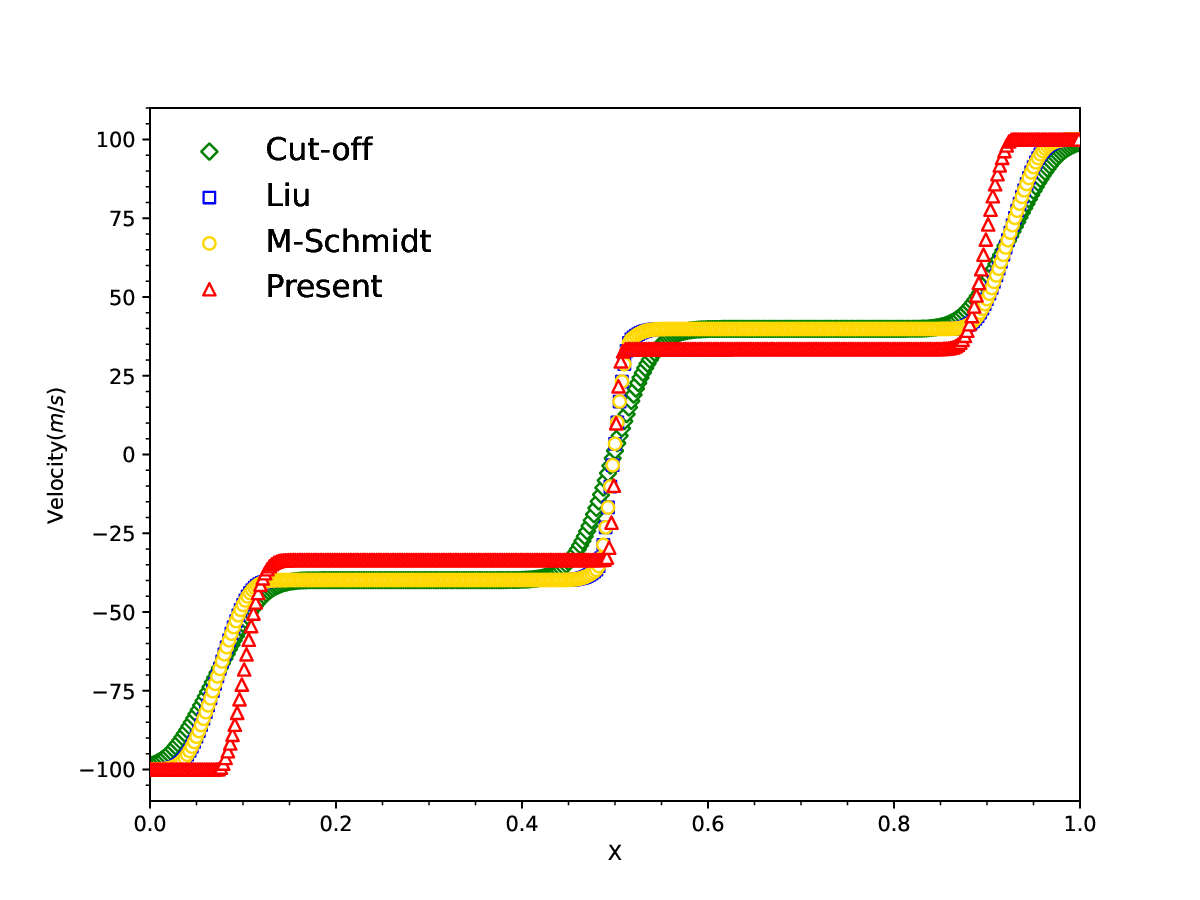}}
    \subfigure[pressure]{\includegraphics[width=0.49\textwidth]{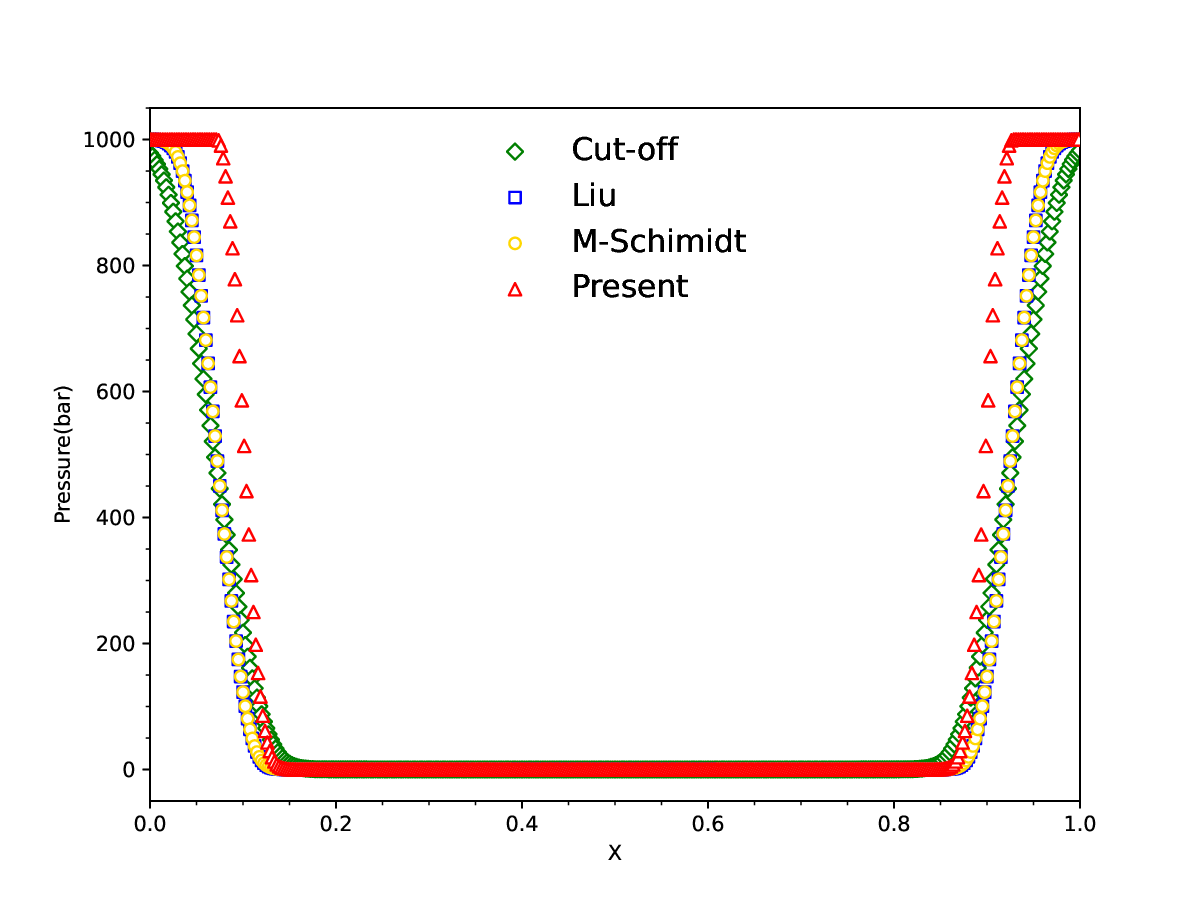}}
    \caption{The comparison of the velocity(left) and pressure(right) profile by the present model to other classic models with the initial velocity magnitude of $100$ m/s at $t = 0.25$ ms in the example \ref{numerical:exmaple1}. }
    \label{numerical:example1:2}
\end{figure}

First, as shown in Fig. \ref{numerical:example1:1}, the numerical velocity results come from the initial velocity magnitude of two flows of 50 m/s at 0.2 ms. Numerical results from the cutoff model, modified Schmidt's model, Liu's isentropic model, and our model are all presented. Fig. \ref{numerical:example1:2} shows the velocity results obtained by our model and the other three models for the initial velocity magnitude is $100$ m/s at $t = 0.25$ ms. It observed that four models provide similar wave structures away from the cavitation region. The cut-off model presents a larger cavitation dimension, while Liu's isentropic model and modified Schmidt's model provide similar but a smaller cavitation dimension. Compared to others, our model shows the smallest dimension of the cavitating region.

\subsubsection{1D cavitating flow in an open tube in one atmosphere}\label{numerical:exmaple2}
Different from example \ref{numerical:exmaple1}, in this part, two water streams move with the velocity of a magnitude of $100$ m/s in the opposite direction from the center of the tube at $x = 0.5$ in one atmosphere. The Courant number is set to 0.4. 

\begin{figure}[t!]
    \centering
    \includegraphics[width=\textwidth]{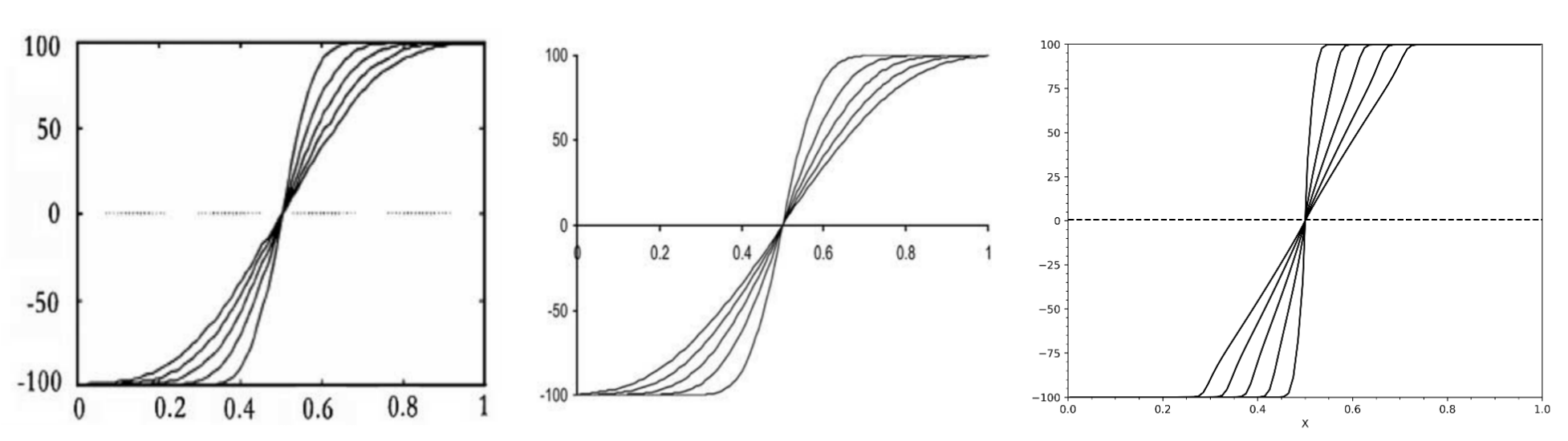}
    \caption{Comparison of velocity profiles at different instant times at 0.5 ms, 1 ms, 1.5 ms, 2.0 ms and 2.5 ms among the Saruel's model(left, duplicated from \cite{Saurel1999multiphase}, Liu's isentropic model(middle, duplicated from \cite{liu2004isentropic}, and present model(right).}
    \label{numerical:example2:1}
\end{figure}

Compared with Saruel and Abgrall's two-phase model\cite{Saurel1999multiphase} and Liu's model\cite{liu2004isentropic}, the numerical results of velocity profiles at five different instant times at 0.5 ms, 1 ms, 1.5 ms, 2.0 ms and 2.5 ms are obtained and shown in Fig. \ref{numerical:example2:1}. We find that velocity profiles in all three models expand along the time evolution. The cavitation region of the present model is the smallest, which is also presented and verified in density and pressure profiles. 

\begin{figure}[ht!]
    \centering
    \subfigure[velocity]{\includegraphics[width=0.49\textwidth]{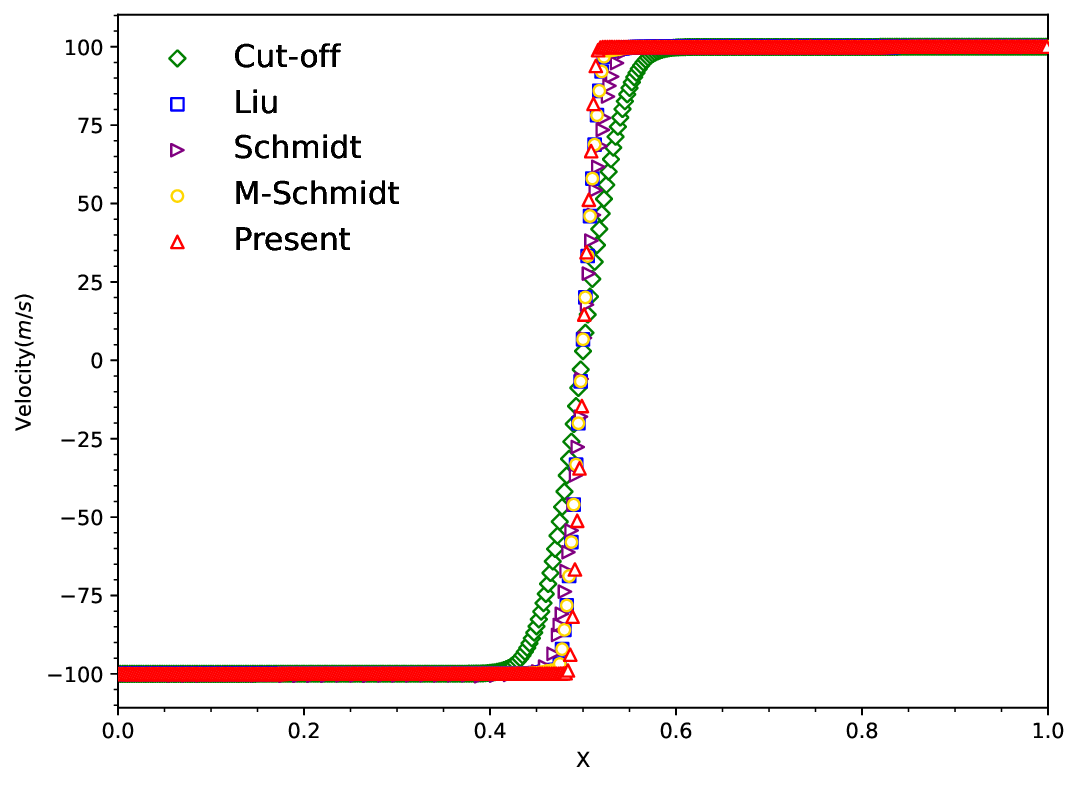}}
    \subfigure[density]{\includegraphics[width=0.49\textwidth]{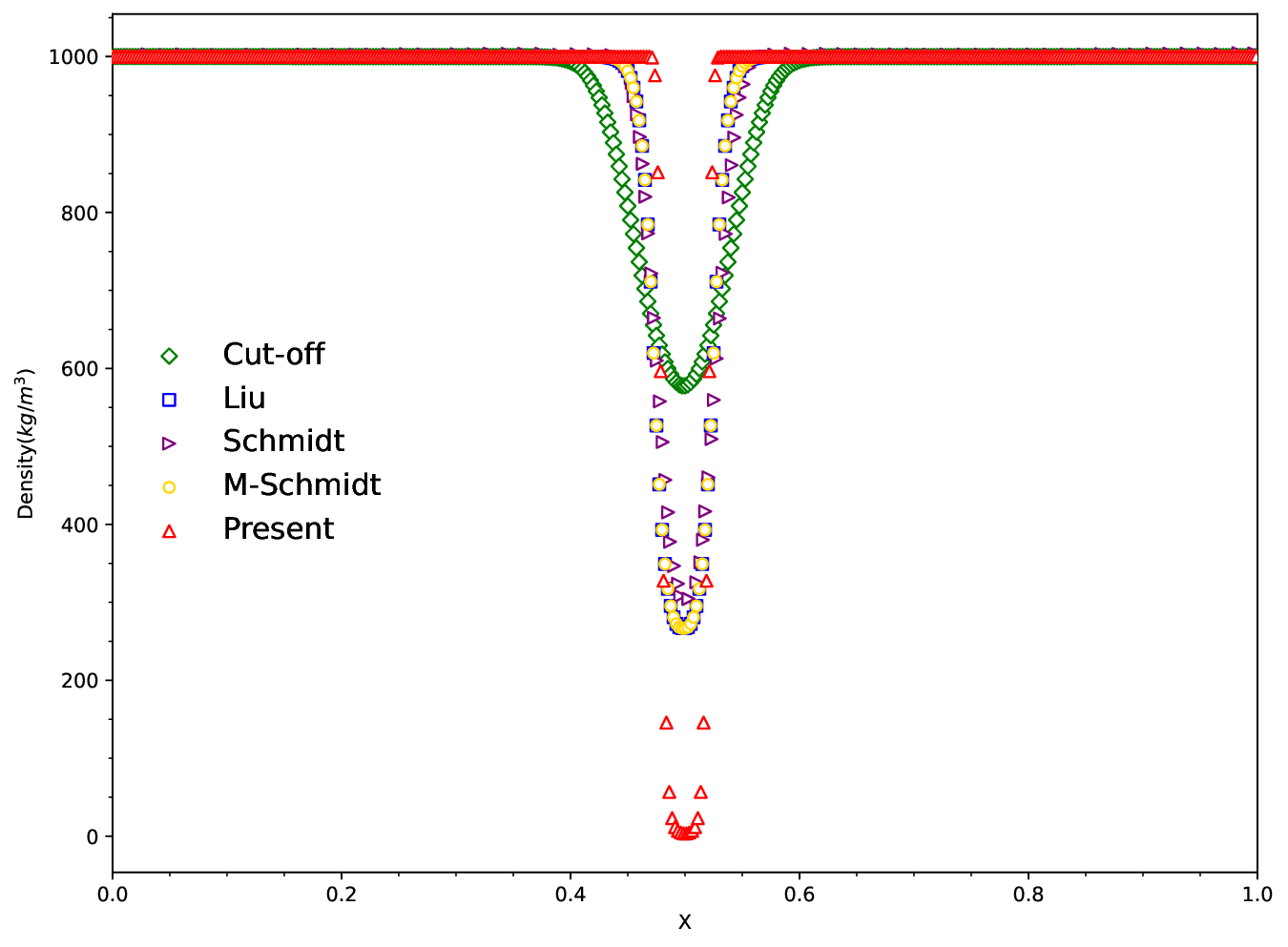}}
    \subfigure[pressure]{\includegraphics[width=0.49\textwidth]{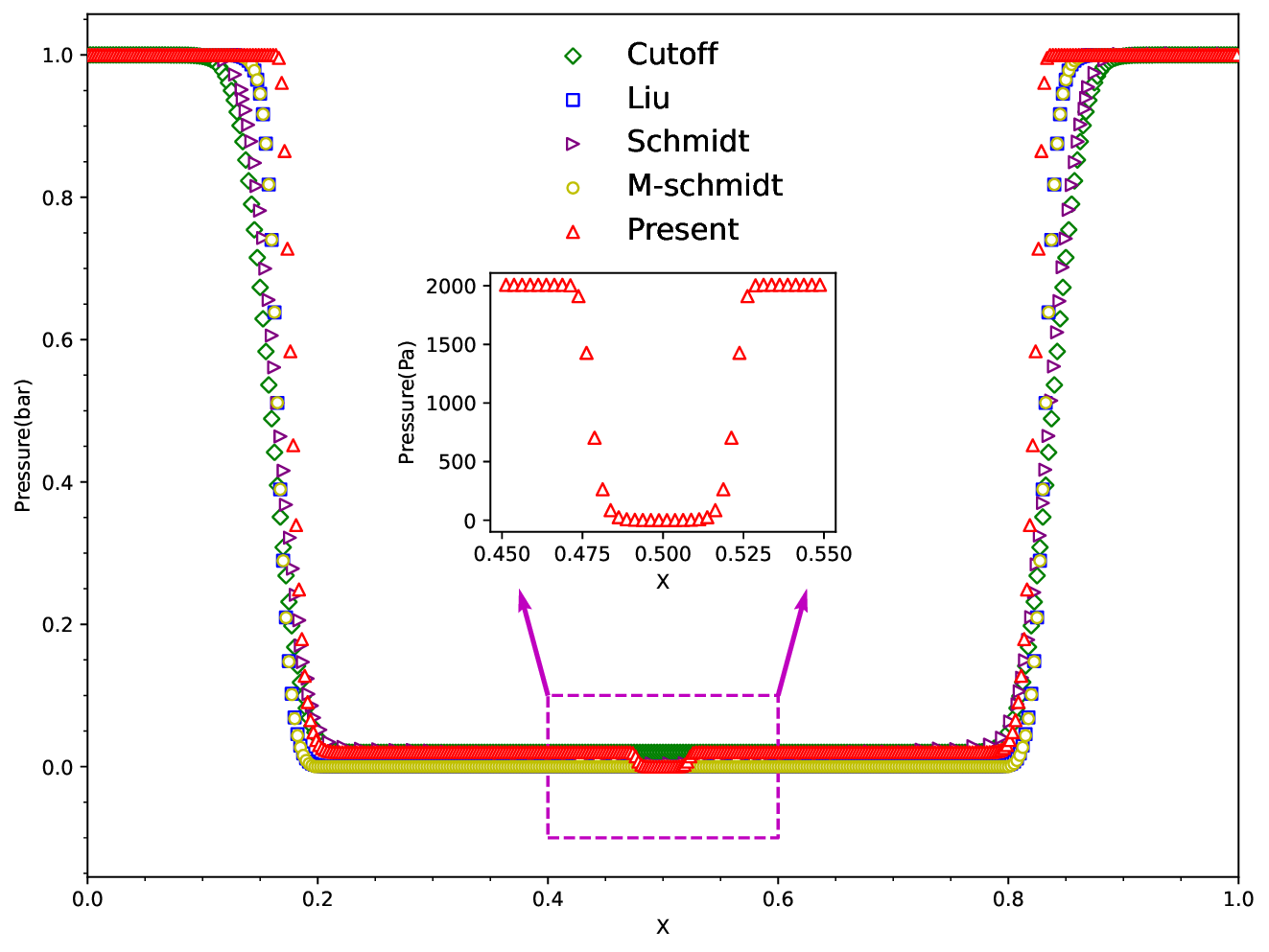}}
    \caption{Numerical results of the velocity, density, and pressure correspond to the cutoff model, Liu's isentropic model, Schmidt's model, modified Schmidt's model, and present model at $t = 0.2$ ms in example \ref{numerical:exmaple2}.}
    \label{numerical:example2::2}
\end{figure}

Fig. \ref{numerical:example2::2} compares the results computed from the cut-off model, Schmidt's model, modified Schmidt's model, Liu's isentropic model and our model, respectively, at time $t = 0.2$ ms. In Fig. \ref{numerical:example2::2}(a), the velocity result among Liu's isentropic model, modified Schmidt's model, and the present model looks very similar. In contrast, the cavitation region produced by Schmidt's model is larger than that of the above three models, while the cut-off model generates twice as large as the others. As shown in Fig. \ref{numerical:example2::2}(b), the density profiles obtained by the five models behave similarly outside the cavitation region. Inside the cavitation region, the cut-off model provides the largest cavitation size. 
However, the lowest density obtained by the cut-off model is nearly double than that of Schmidt's model, modified Schmidt's model, and Liu's isentropic model and much higher than our model. Although the lowest density profile in Liu's isentropic model and other two Schmit-type models are almost half of the cut-off model, it can not reach nearly zero while the pressure is nearly zero as shown in Fig. \ref{numerical:example2::2}(c), which is not physical. According to experimental data, the density tends to zero when the pressure tends to zero. In our model, we find that the cavitation size is the smallest, nearly half of the cut-off model. The lowest density in our model is nearly zero, which is more physical and reasonable. The pressure distribution is presented in Fig. \ref{numerical:example2::2}(c). The pressure jump across the cavitation boundary ranges from 0.39 to 0.61 in Liu's isentropic model, 0.41 to 0.59 in Schmidt's model while ranges from 0.31 to 0.69 in the cut-off model \cite{liu2004isentropic}. As shown in the embedded small figure in Fig. \ref{numerical:example2::2}(c), the pressure jump in our model across the cavitation boundary ranges from 0.46 to 0.54, which is the smallest size of the cavitation region, as demonstrated before. 

\subsubsection{1D cavitating flow in a closed tube}\label{numerical:exmaple3}
This numerical example is to test the shock-cavitation interaction \cite{liu2004isentropic}. This case resembles example \ref{numerical:exmaple2} while the boundary condition is considered reflective. The Courant number of this test is set to 0.4. As time evolves, a shock will be created at each end of the tube and move towards the center due to the rigid wall boundary condition. The flow is initially pure water and transforms into a vapor-water mixture at the center of the tube. 

\begin{figure}[ht!]
    \centering
    \subfigure[density]{\includegraphics[width=0.49\textwidth]{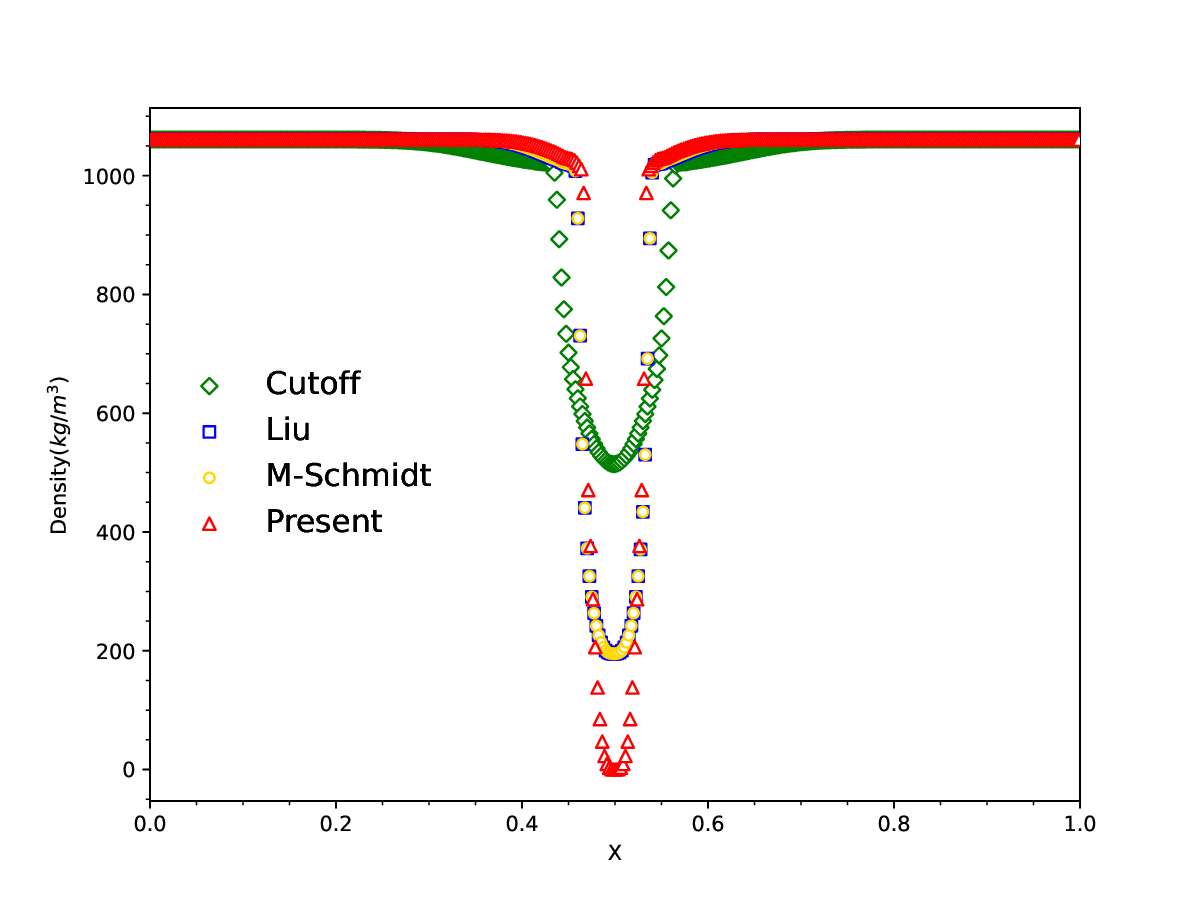}}
    \subfigure[pressure]{\includegraphics[width=0.49\textwidth]{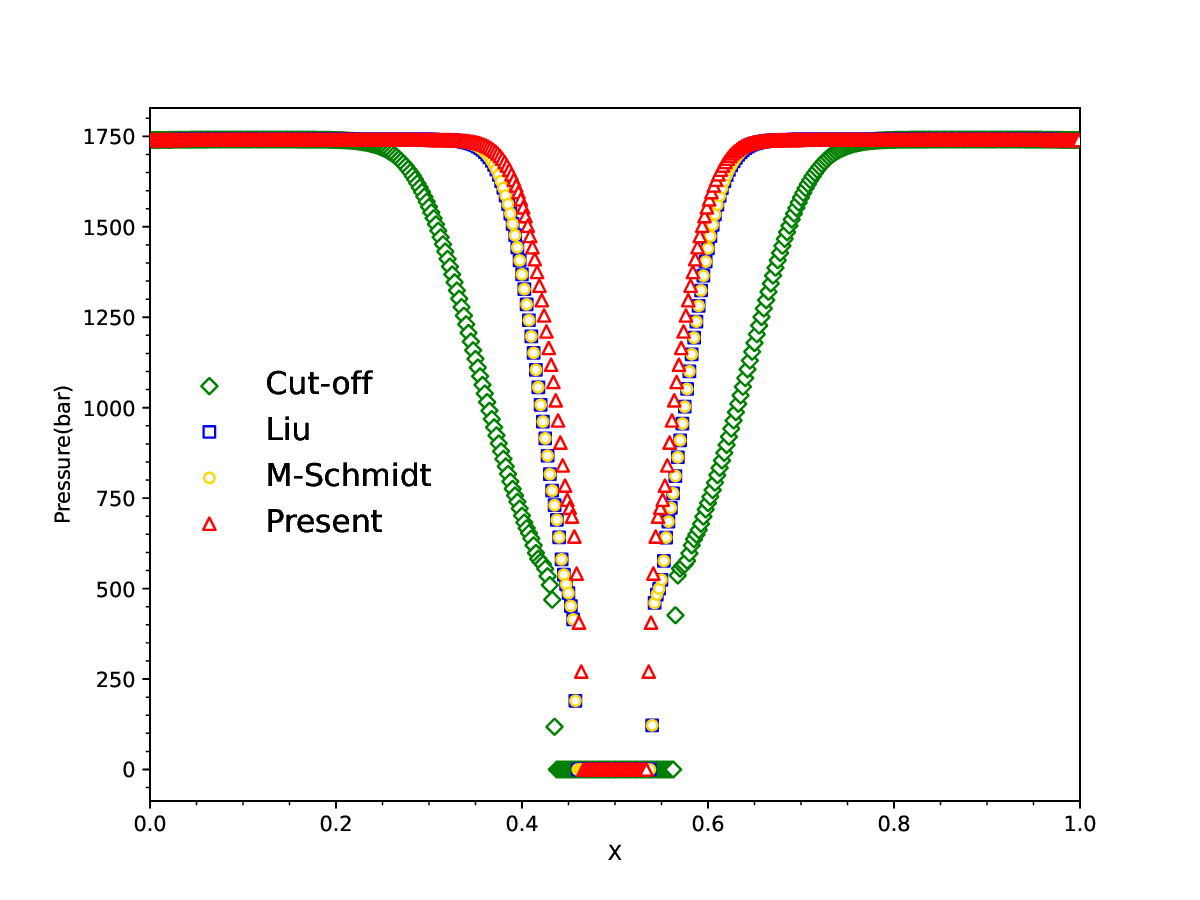}}
    \caption{Numerical results of the velocity, density, and pressure correspond to the cutoff model, Liu's isentropic model, modified Schmidt's model, and the present model at $t = 0.3$ ms in example \ref{numerical:exmaple3}.}
    \label{numerical:example3:1}
\end{figure}
Fig. \ref{numerical:example3:1} shows the numerical results of the cut-off model, Liu's isentropic model, modified Schmidt's model, and our model, at the time $t = 0.3$ ms. The results cut-off model behaves differently from the other three models. The cut-off model shows the highest value of lowest density and a cavitation region which is twice larger than the other models. Liu's isentropic model and modified Schmidt's model behave similarly in both density and pressure profiles. The lowest density in Liu's isentropic model is nearly half of the cut-off model. However, it does not tend to approach zero. Our model gives more reasonable density profiles during the low-pressure region, which almost approaches zero. It is also observed that two shocks created at each end of the tube move to the center and then meet the rarefaction generated at the center. Moreover, the cut-off model results in the largest cavitation region, while our model leads to the narrowest one.


\subsection{Two-dimensional problems}
In this part, we present three two-dimensional problems in engineering applications, which are carried out on quadrangle unstructured meshes for each phase of fluids. In these problems, parallel computing based on the classical domain decomposition methods and the $h$-adaptive mesh method are implemented to improve the efficiency of the simulation.

\subsubsection{Nuclear underwater blast problem}
\begin{figure}[ht!]
    \centering
    \subfigure[Peak overpressure]
    {\includegraphics[width=0.45\textwidth]{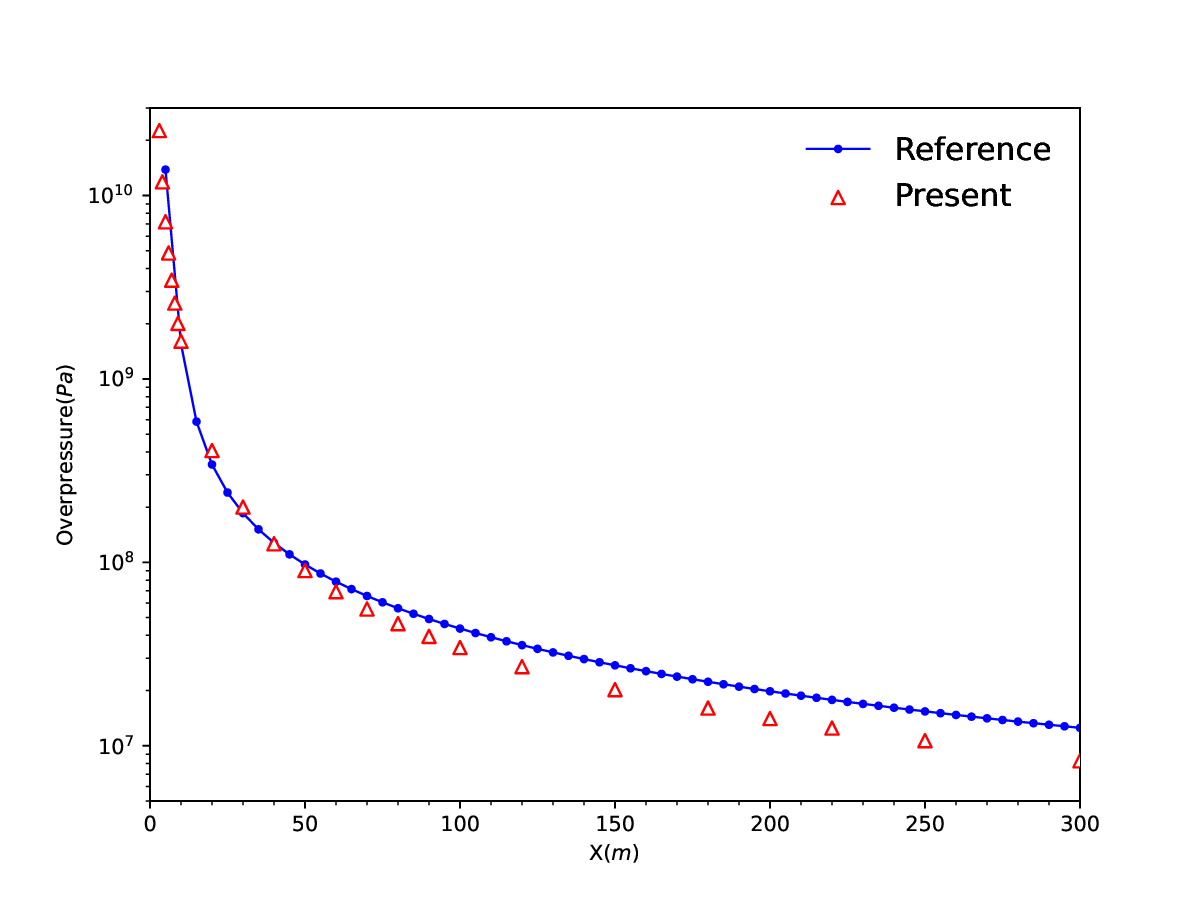}} ~~
    \subfigure[Impulse]
    {\includegraphics[width=0.45\textwidth]{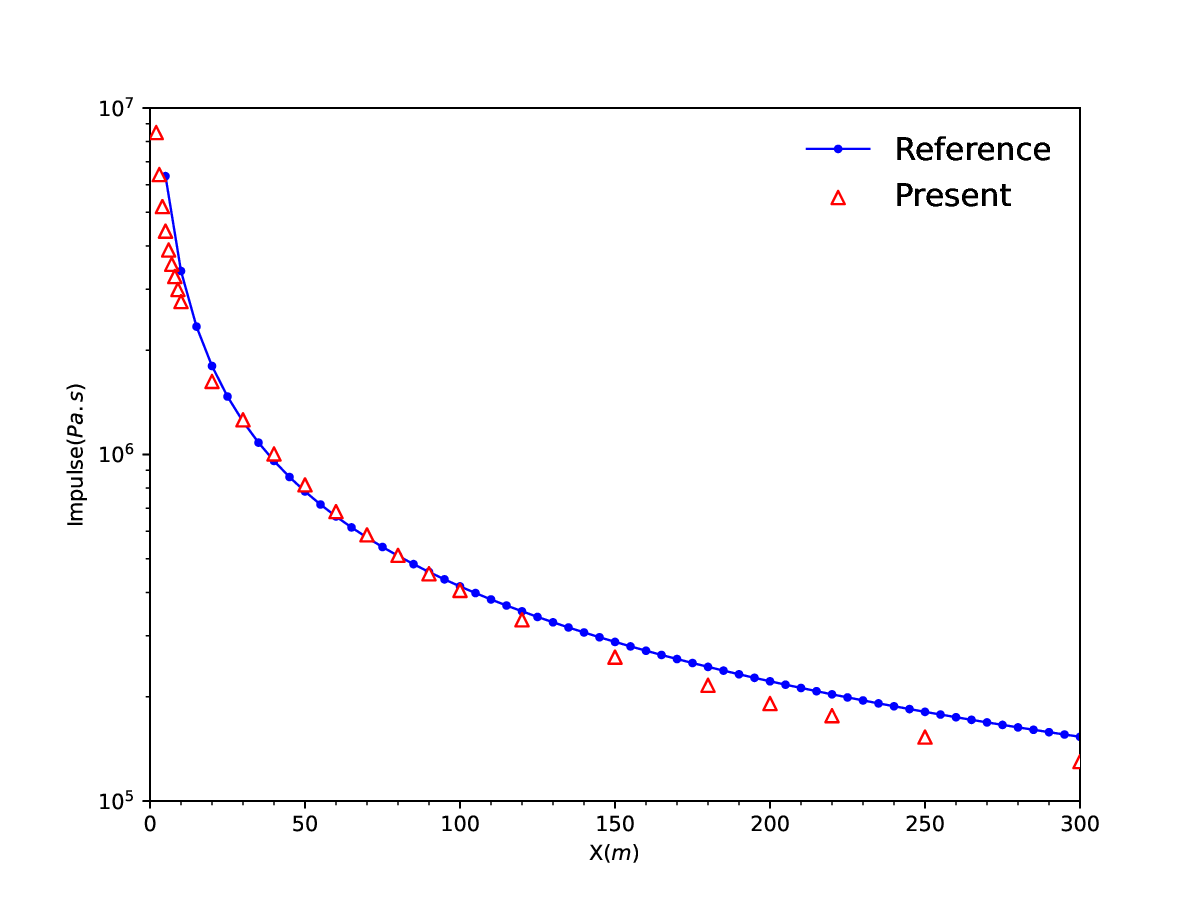}}
    \caption{Shock wave parameters for nuclear underwater blast problem.}
    \label{rm::nuc_blast_ref}
\end{figure}
\begin{figure}[hb!]
\centering
\subfigure[$t=0.01$]{
\includegraphics[width=0.42\textwidth]{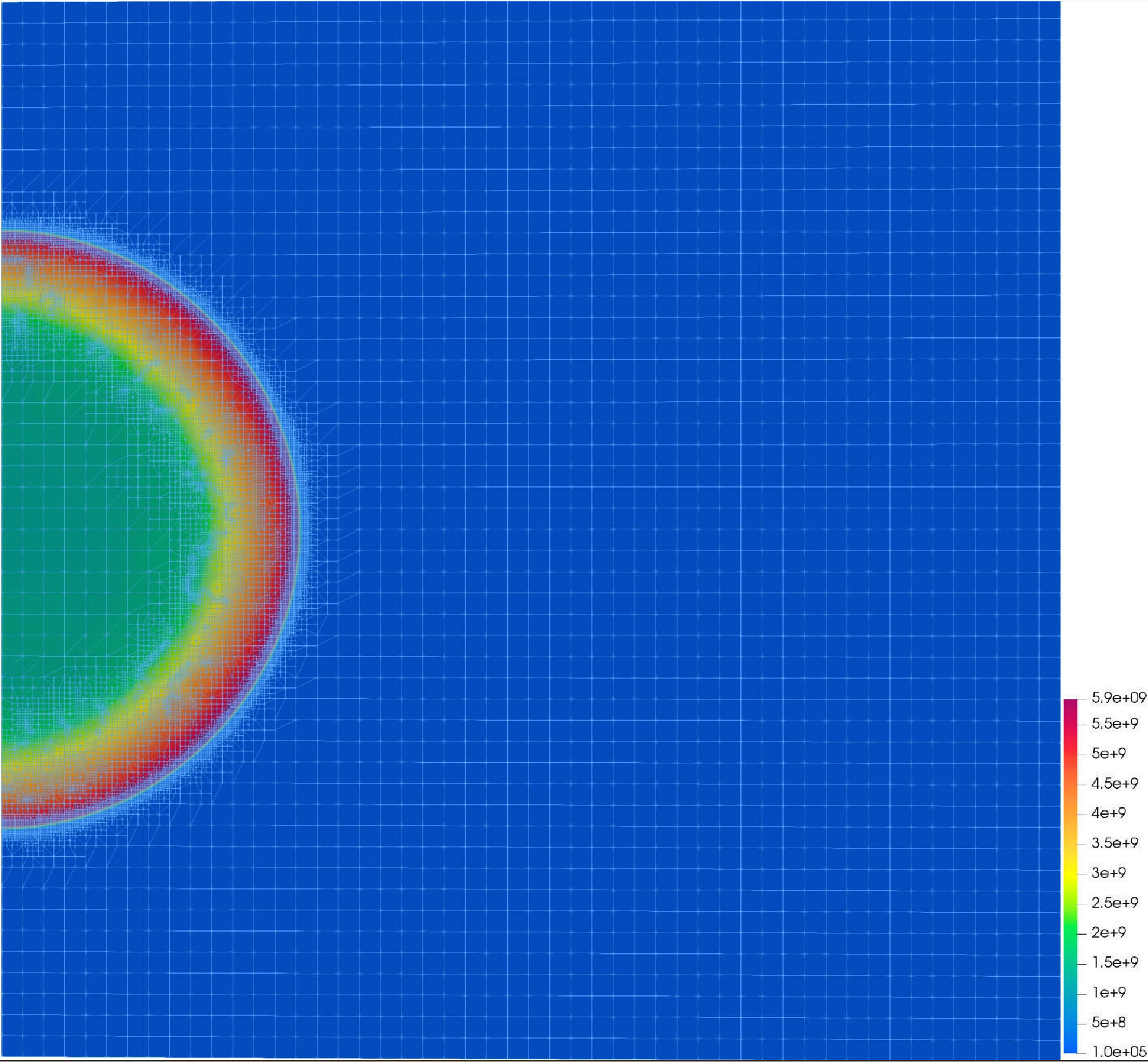}}
\subfigure[$t=0.02$s]
{\includegraphics[width=0.42\textwidth]{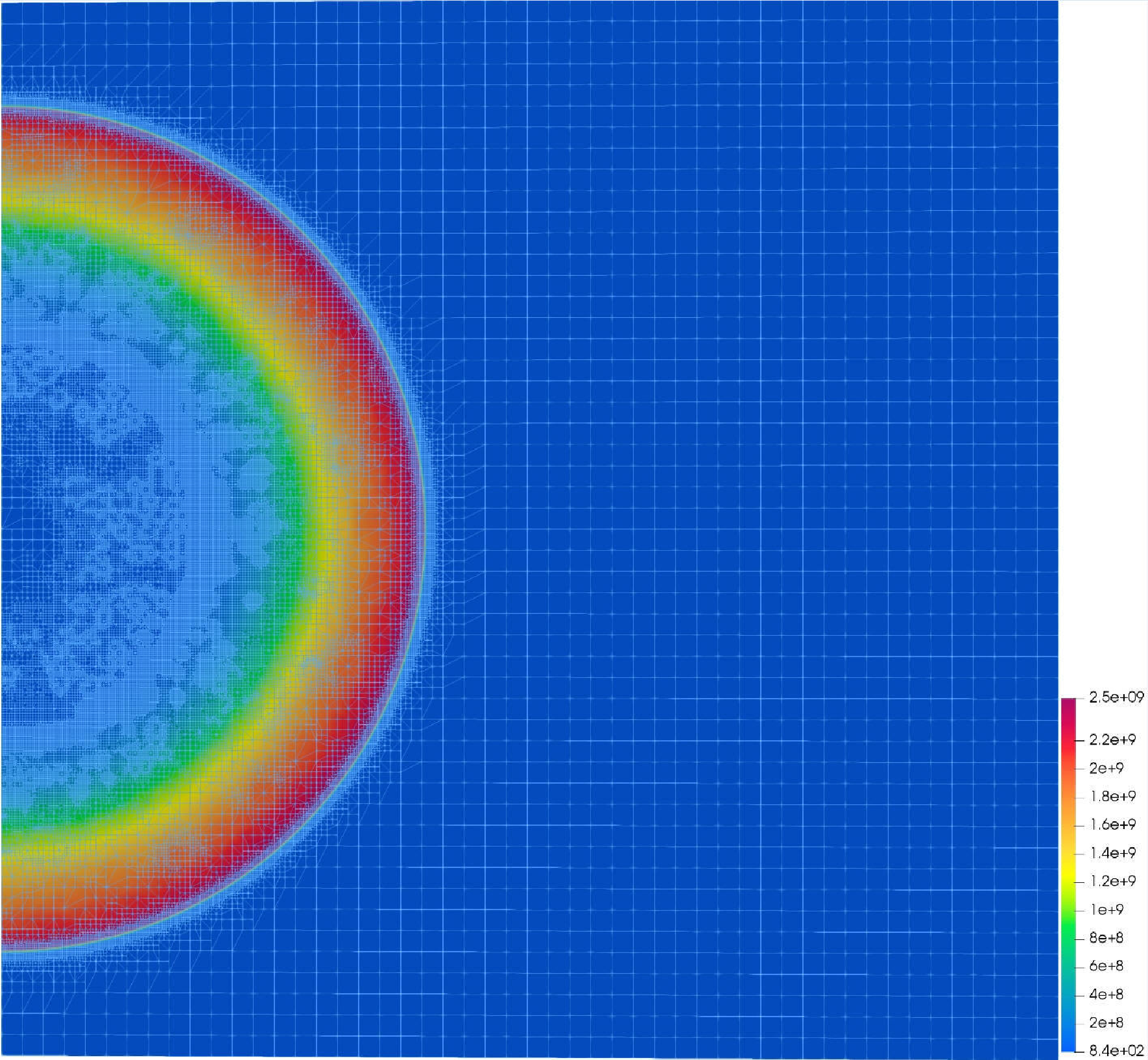}}
\caption{Shock wave contours of the nuclear underwater blast.}
\label{rm::nuc_blast}
\end{figure}

In this example, we simulate a nuclear underwater blast problem in a computational domain $0\le r,z\le 300$ m, within the cylindrically symmetric coordinate system, which has the following governing equations
\[
\dfrac{\partial}{\partial t}
\begin{bmatrix}
r\rho \\ r\rho u \\ r\rho v\\ rE
\end{bmatrix}
+\dfrac{\partial}{\partial r}
\begin{bmatrix}
r\rho u \\ r(\rho u^2+p) \\ r\rho uv \\ r(E+p)u
\end{bmatrix}
+\dfrac{\partial}{\partial z}
\begin{bmatrix}
r\rho v \\ r\rho uv \\ r(\rho v^2+p)\\ r(E+p)v
\end{bmatrix}
=\begin{bmatrix}
0 \\ p \\ 0 \\ 0
\end{bmatrix}.
\]
The whole boundaries are set as the outflow boundary conditions except that the left edge $r=0$ and the bottom edge $z=0$ are set as rigid walls. The equation of state of the nuclear explosion products is described by the dnnEOS developed in \cite{Li2023application}, and the water is described by the polynomial EOS and cavitation model whose parameters have the same values as the previous examples. The explosive center is located at $r=0,~z=150$ m, and the radius of the phase interface is 0.3 m at $t=0$. Fig. \ref{rm::nuc_blast} shows the pressure contours and adaptive meshes at $t=0.01, ~0.02,~0.04$ and $0.08$ s. The peak overpressure and impulse at different radii are shown in Fig. \ref{rm::nuc_blast_ref}. Our numerical results agree well with the reference data interpolated from the given standard data in \cite{Glasstone1977}.
\begin{figure}[ht!]
\centering
\subfigure[$t=0.04$s]
{\includegraphics[width=0.42\textwidth]{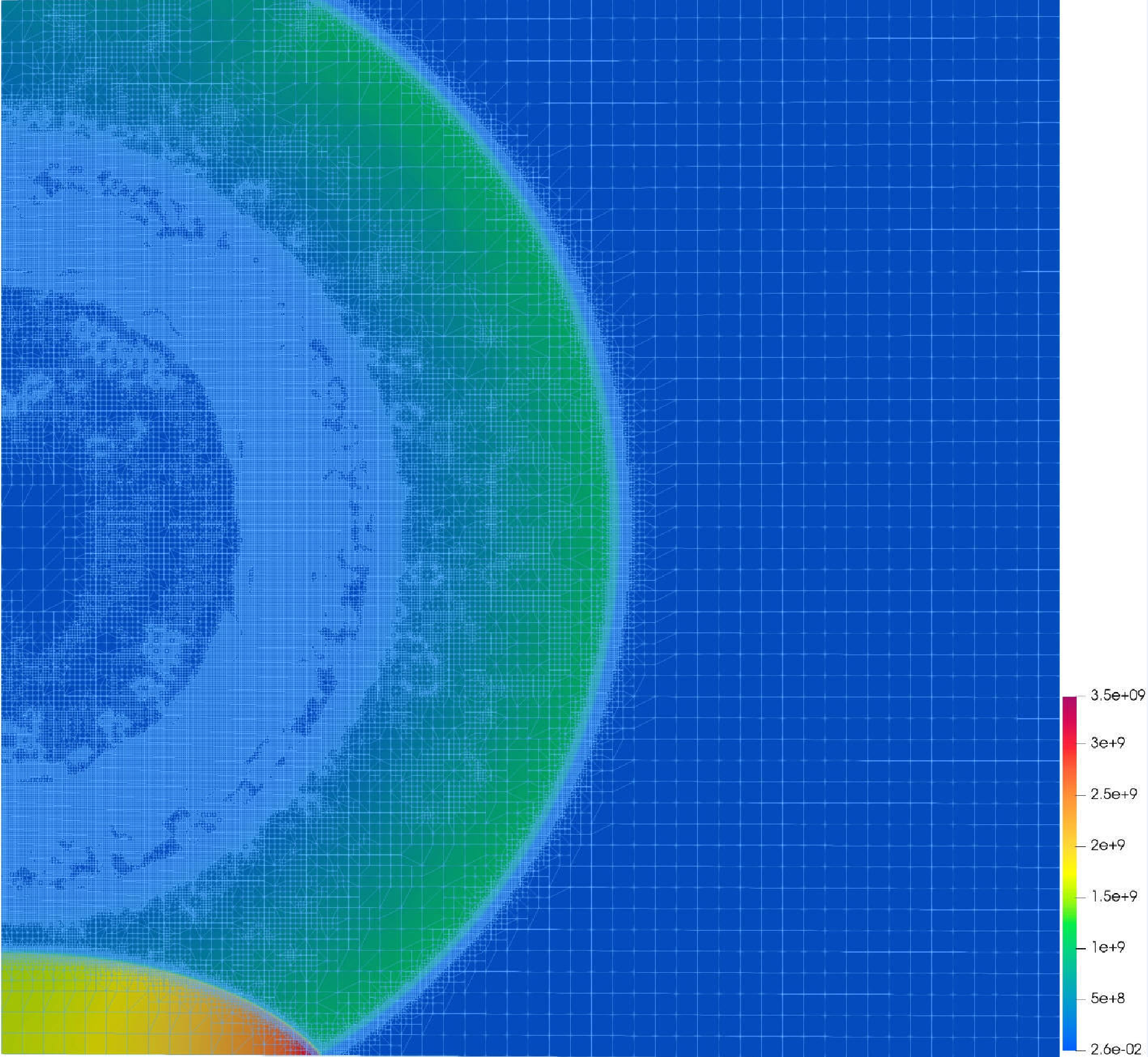}}
\subfigure[$t=0.08$s]
{\includegraphics[width=0.42\textwidth]{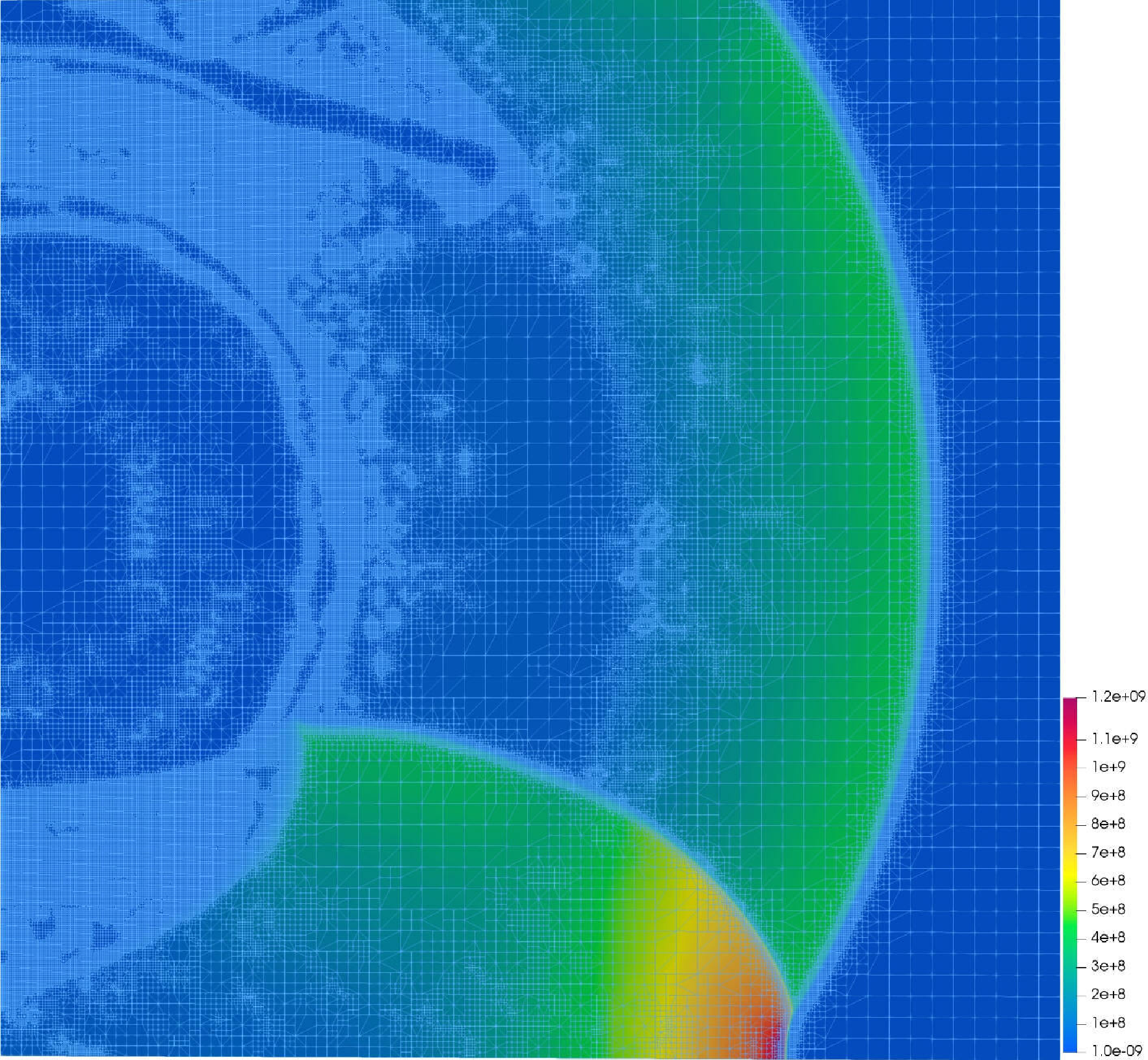}}  
\caption{Continuity of \ref{rm::nuc_blast}}
\end{figure}
\begin{figure}[ht!]
    \centering
    \begin{minipage}[t]{0.49\textwidth}
        \centering
        \includegraphics[width=0.5\textwidth,height = 0.5\textwidth]{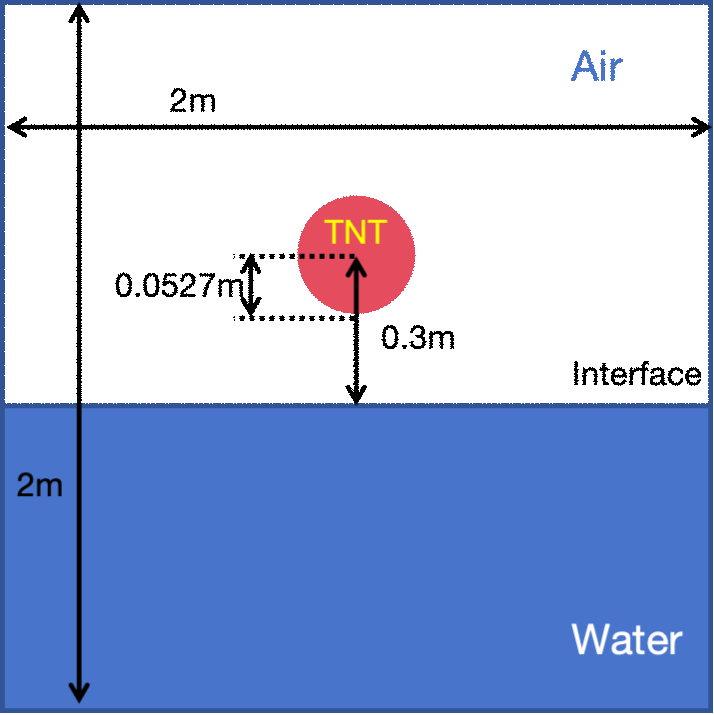}
        \caption{Model of TNT explosion above the gas-water surface.}
        \label{rm:ns1}
    \end{minipage}
    \begin{minipage}[t]{0.49\textwidth}
        \centering
        \includegraphics[width=1.0\textwidth, height = 0.5\textwidth]{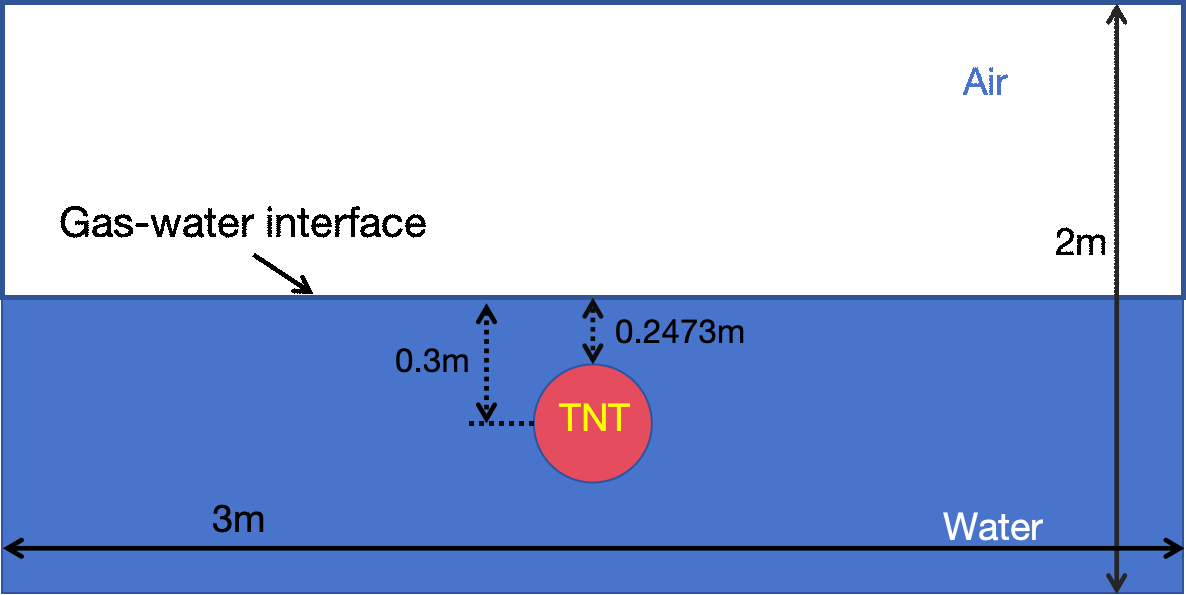}
        \caption{Model of TNT explosion underwater near the free surface.}
        \label{rm:ns2}
    \end{minipage}
\end{figure}

\subsubsection{TNT explosion above gas-water surface}\label{subsec:near2d}
In this problem, we simulate a TNT air explosion nearly above the gas-water interface, shown in Fig. \ref{rm:ns1}. The computational domain is $[0,2]$ m $\times[-1,1]$ m in meters. The explosive center is located at $(x,y)=(0,0.3)$ m with an initial radius of $0.0527$ m at $t=0$. The region $y\le 0$ is filled with water, while the remaining is filled with air. The initial conditions are 
\[
[\rho, u, v, p]^\top = \left\{
\begin{array}{ll}
[1630, ~0, ~0, ~9.5 \times 10^{9}]^\top, & \sqrt{(x-1)^2+(y-0.3)^2} \le 0.0527,\\ [1mm]
[1.29, ~0, ~0, ~1.0\times 10^5]^\top, & \sqrt{(x-1)^2+(y-0.3)^2} > 0.0527 \text{ and } y > 0, \\ [1mm]
[1000, ~0, ~0, ~1.0\times 10^5]^\top, & y \le 0.
\end{array}
\right.
\]

The interaction between the TNT product, air, and water is described by the five-equation model \eqref{model:five_equation}, a diffuse interface method to simulate the multiphase flows. We take the JWL EOS and polynomial EOS to simulate the TNT product and water, respectively, whose parameters have the same values as that in Section \ref{sec::underwater_sphere}, and the air is described by the ideal gas EOS with $\gamma=1.4$.

\begin{figure}[t!]
    \centering
    \subfigure[pressure contours at $t=3\times10^{-5}$]{\includegraphics[width=0.45\textwidth]{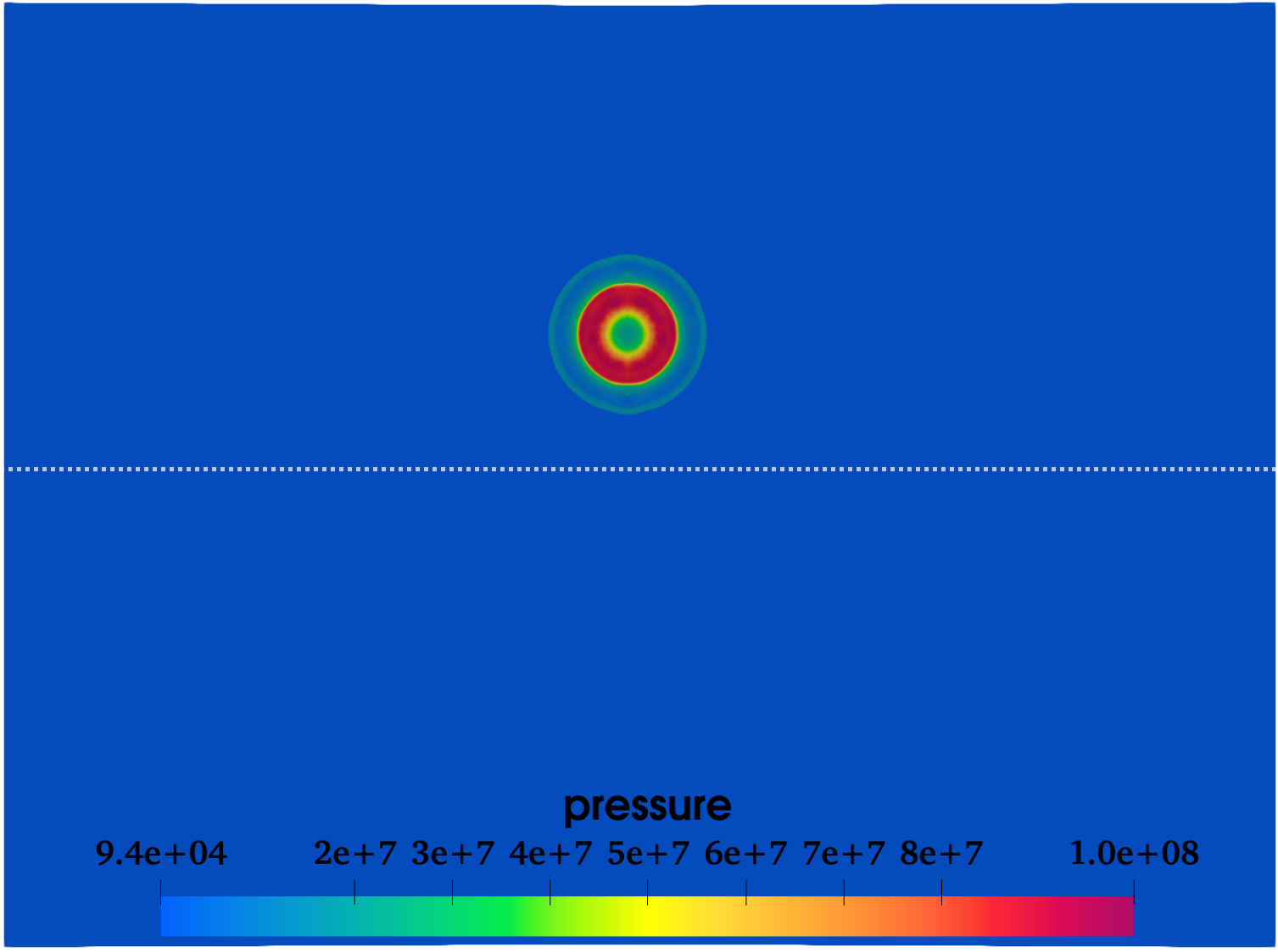}}
    \subfigure[adaptive mesh at $t=3\times10^{-5}$]{\includegraphics[width=0.45\textwidth]{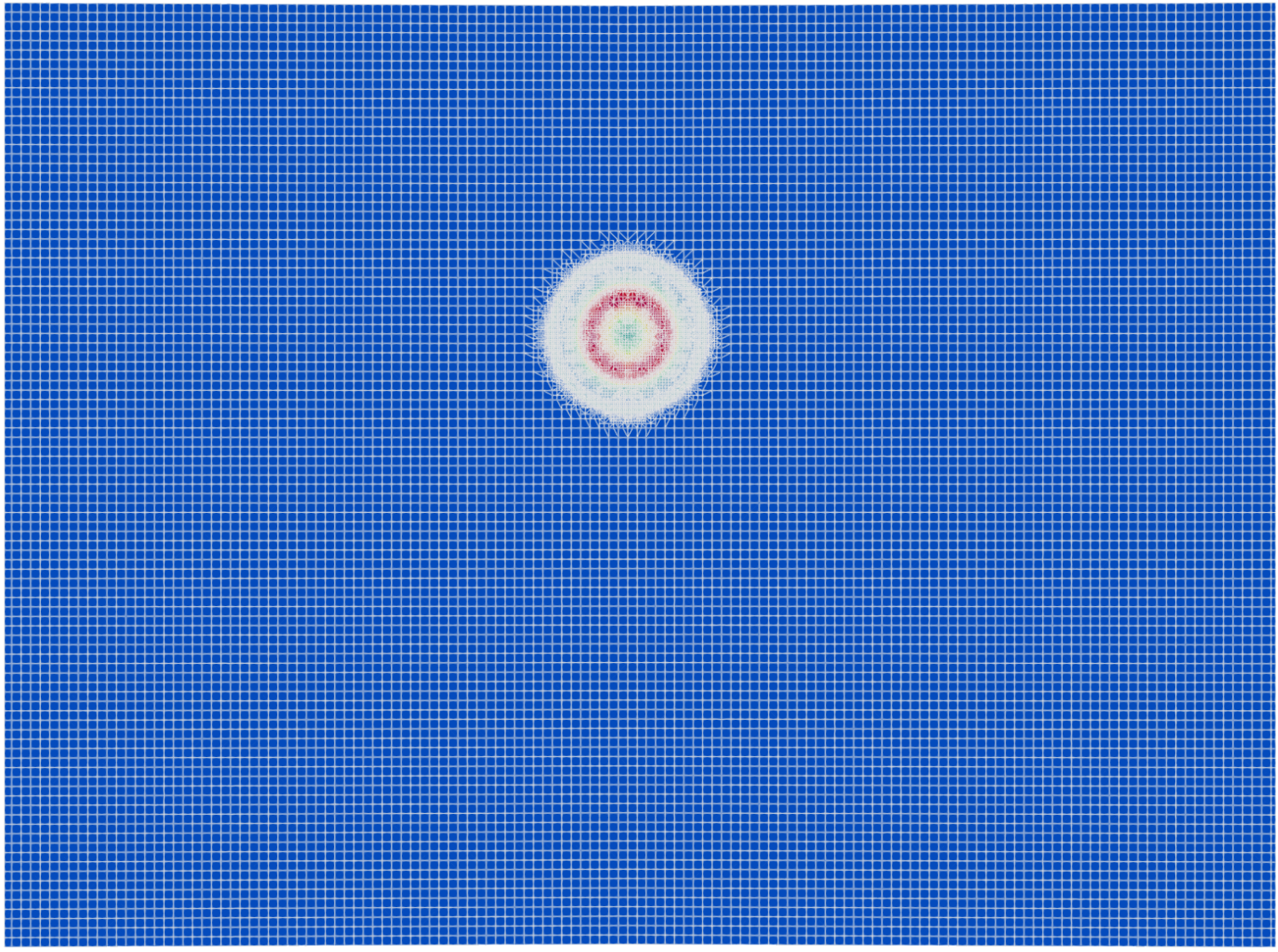}} \\
    \subfigure[pressure contours at $t=6\times10^{-5}$]{\includegraphics[width=0.45\textwidth]{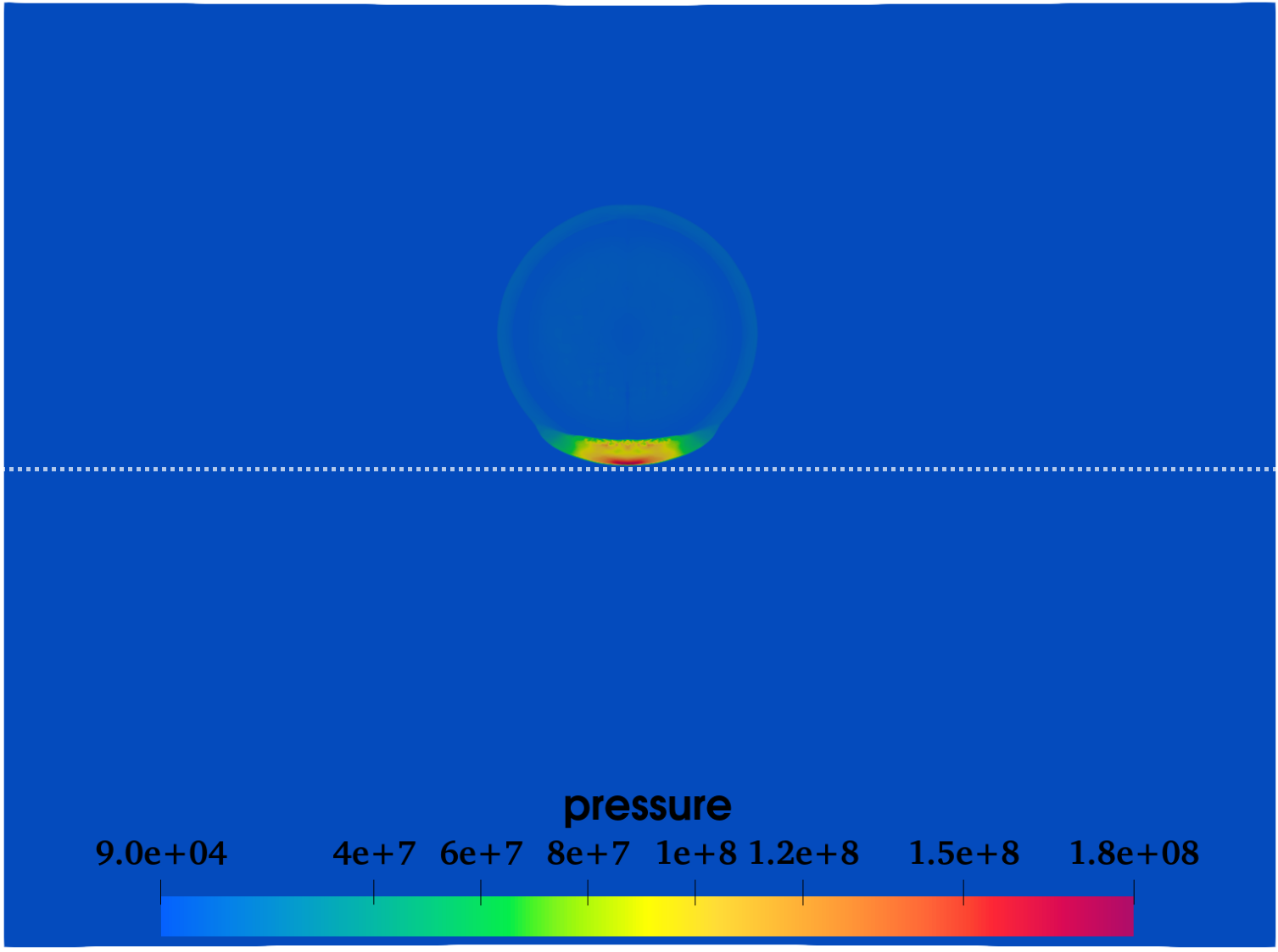}}
    \subfigure[adaptive mesh at $t=6\times10^{-5}$]{\includegraphics[width=0.45\textwidth]{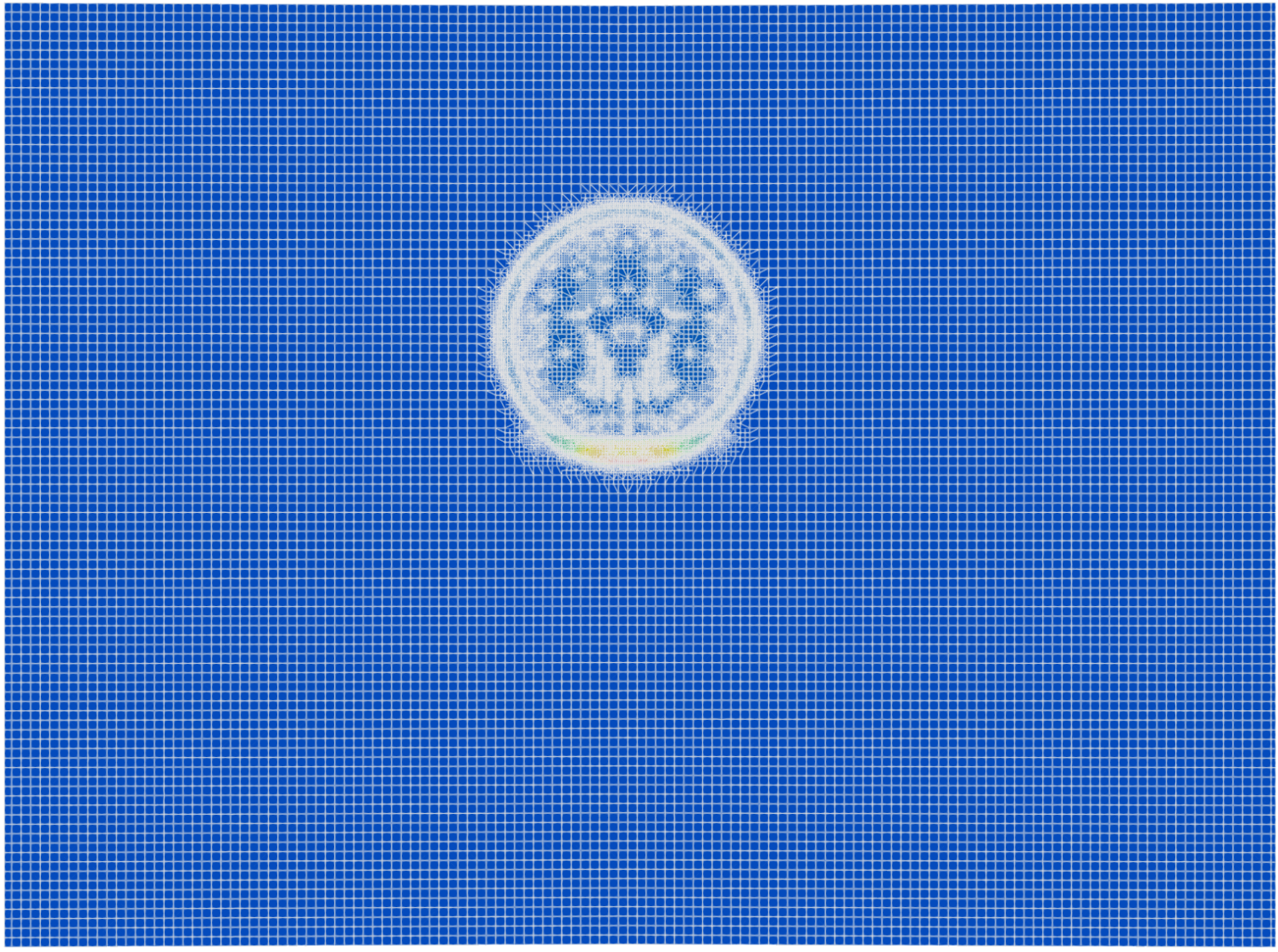}}\\
    \subfigure[pressure contours at $t=4\times10^{-4}$]{\includegraphics[width=0.45\textwidth]{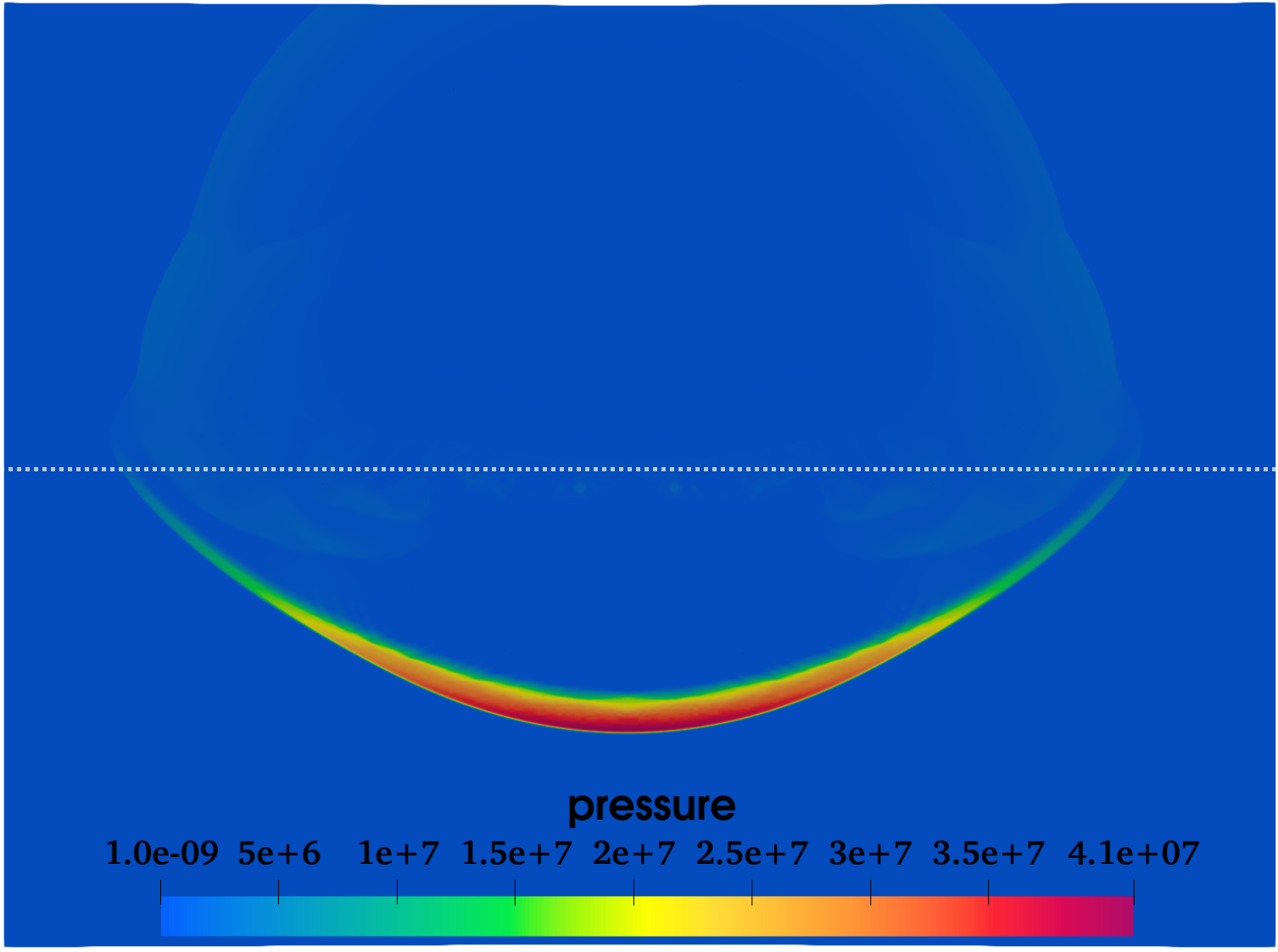}}
    \subfigure[adaptive mesh at $t=4\times10^{-4}$]{\includegraphics[width=0.45\textwidth]{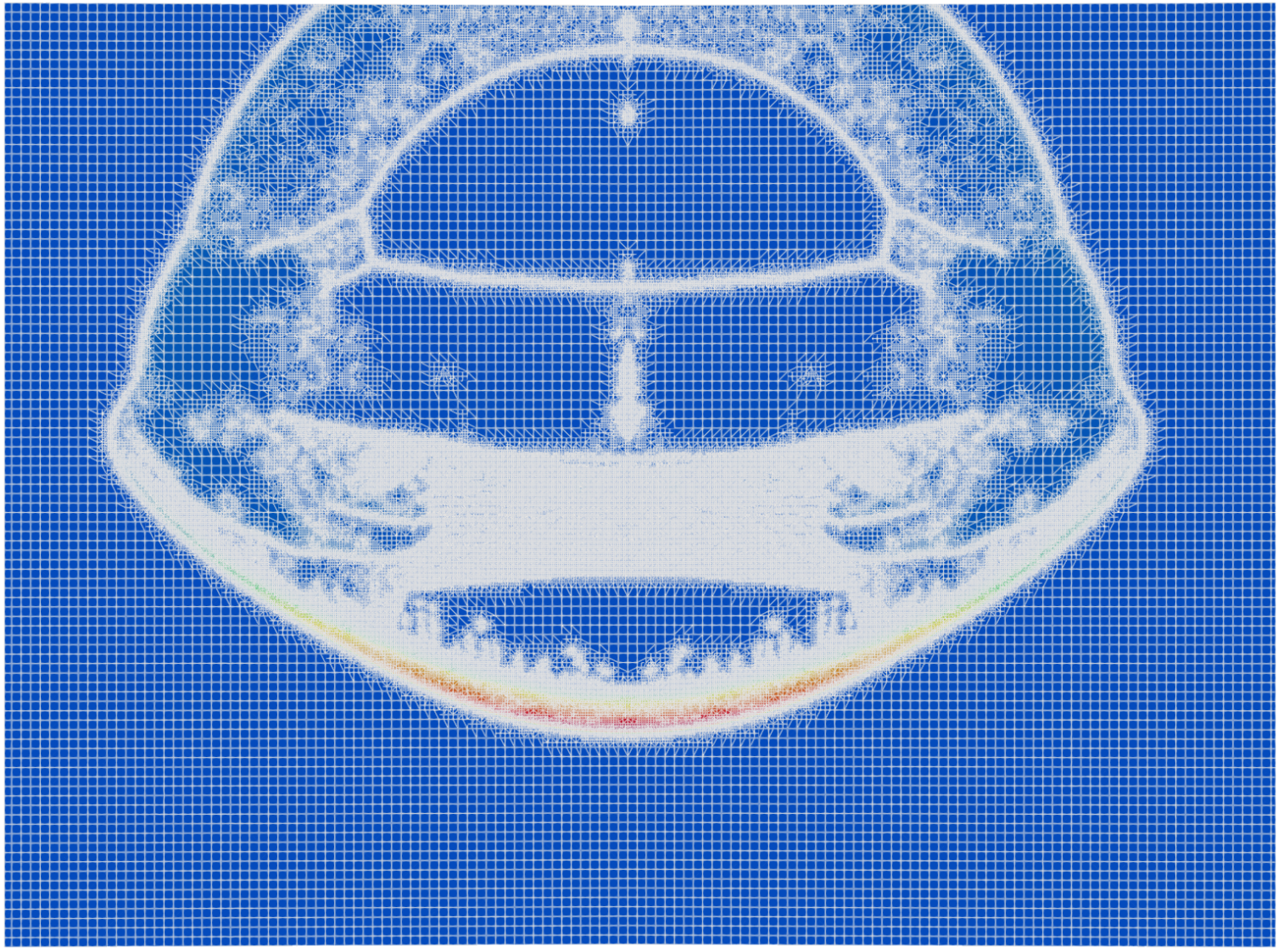}}\\
    \caption{Shock wave contours of TNT explosion near gas-water surface.}
    \label{rm:near}
\end{figure}

Fig. \ref{rm:near} shows the pressure contours of the whole computational domain at typical times, including the TNT explosive, air, and water. When the blast wave produced by the TNT product arrives at the gas-water interface, part of the blast wave is reflected and propagates along the gas-water interface simultaneously. As time increases, the reflective wave will transform from regular reflection to irregular Mach reflection.

Additionally, the remaining blast wave will transmit into the water and propagate downwards. Due to the great discrepancy in the wave impedance between the water and air, the transmitted wave is stronger than the reflected one. From Fig. \ref{rm:near}, we can see that our numerical scheme can correctly simulate the blast wave propagation in the air and water.


\subsubsection{Underwater explosion near free surface}
\label{rm:underwater2dimensioanl}
In this example, we simulate a spherical underwater explosion near the free surface, shown in Fig. \ref{rm:ns2}, where a spherical TNT explosion product is located below the gas-water interface with a radius of $r=0.1$ m. The whole computational domain is $[0,3]$ m $\times[-1,1]$ m. The location of the explosion point is set as $x=0,y=-0.3$ m. We take the JWL EOS, polynomial EOS and ideal gas EOS to simulate the TNT, water, and air, respectively. The initial conditions are
\[
 [\rho, u, v, p]^\top = \left\{
    \begin{array}{ll}
      [1630, ~0, ~0, ~9.5\times 10^{9}]^\top, & x^2+(y+0.3)^2 \le 0.05,            \\ [2mm]
      [1000, ~0, ~0, ~1.0\times 10^{5}]^\top, & x^2+(y+0.3)^2 > 0.05 \text{ and } y \le 0, \\ [2mm]
      [1.0, ~0, ~0, ~1.0\times 10^{5}]^\top,  & y > 0.
    \end{array}
  \right.
\] 

\begin{figure}[htbp]
\centering
\subfigure[pressure contours at $t=100 \mu$s]
{\includegraphics[width=0.42\textwidth]{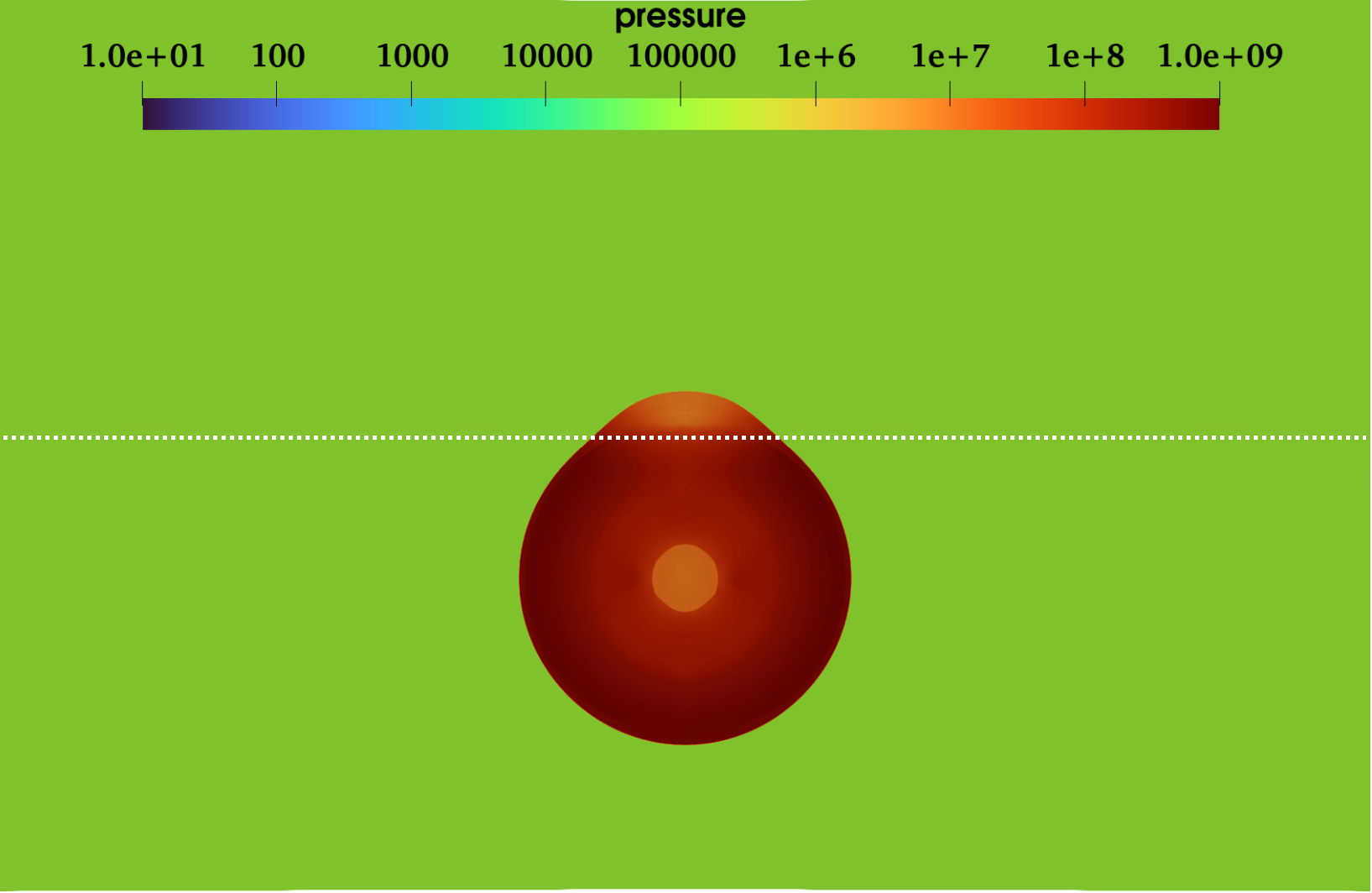}} ~~
\subfigure[pressure contours at $t=210 \mu$s]
{\includegraphics[width=0.42\textwidth]{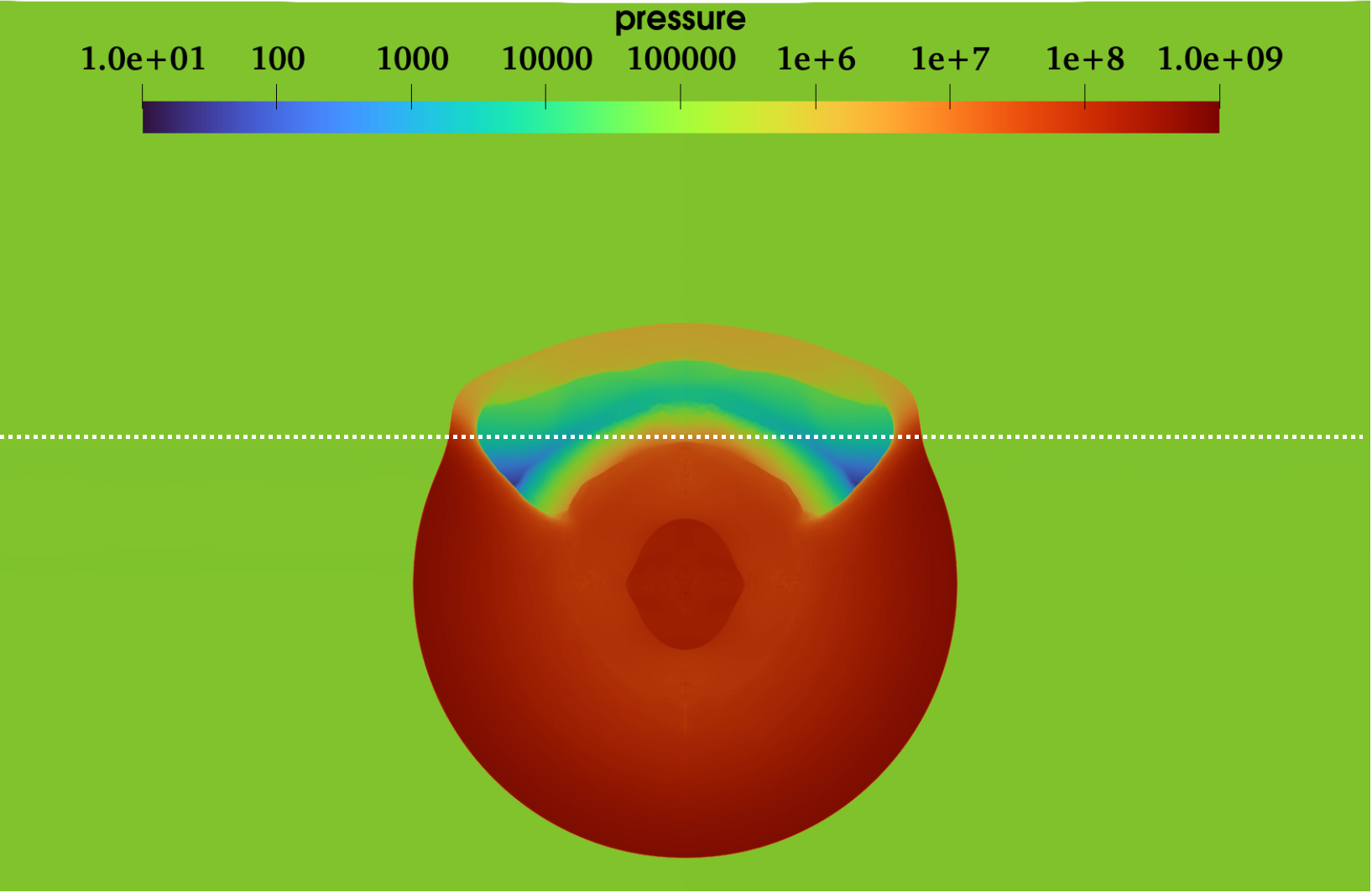}}  \\
\subfigure[pressure contours at $t=260 \mu$s]
{\includegraphics[width=0.42\textwidth]{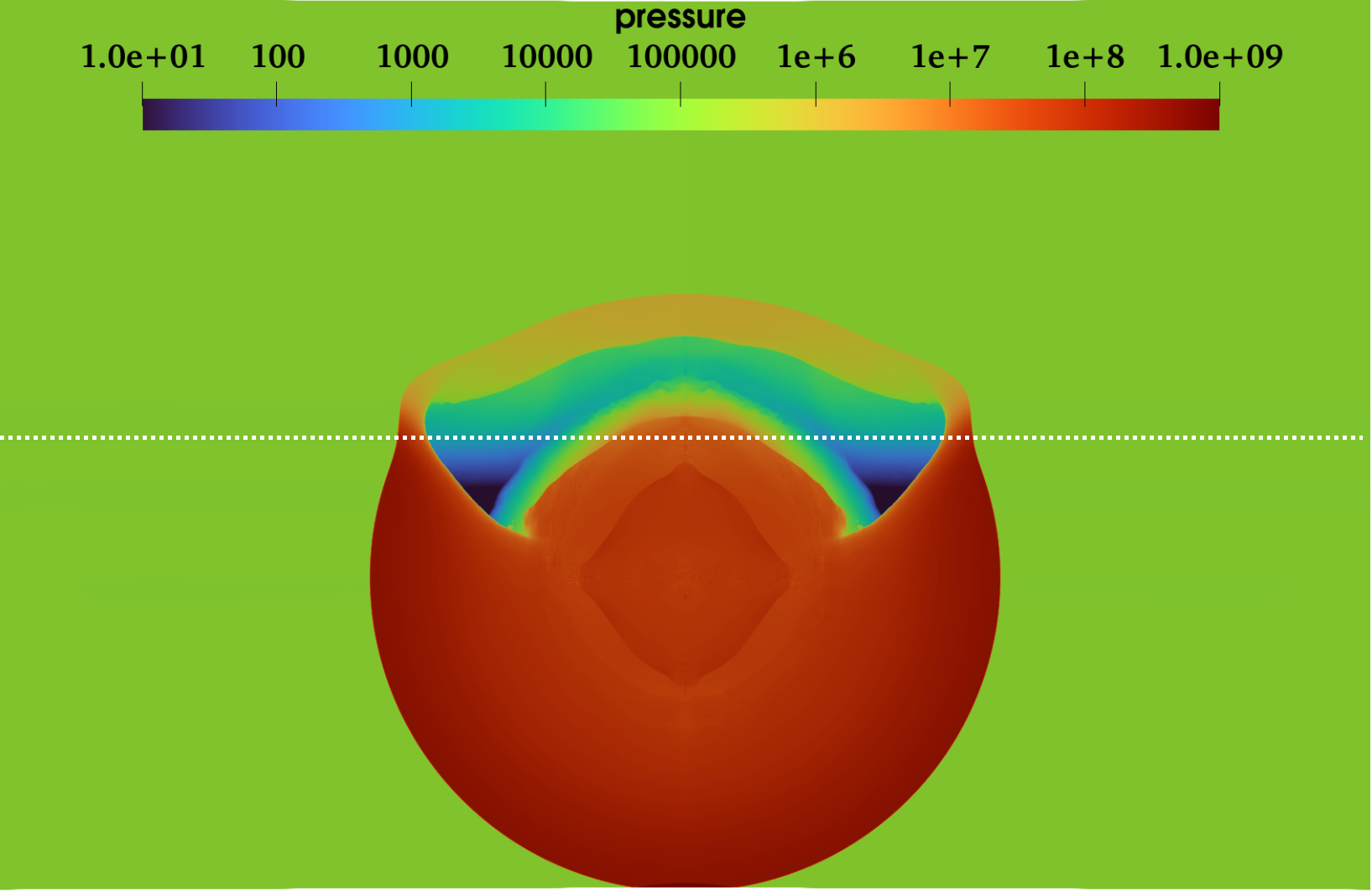}} ~~
\subfigure[pressure contours at $t=315 \mu$s]
{\includegraphics[width=0.42\textwidth]{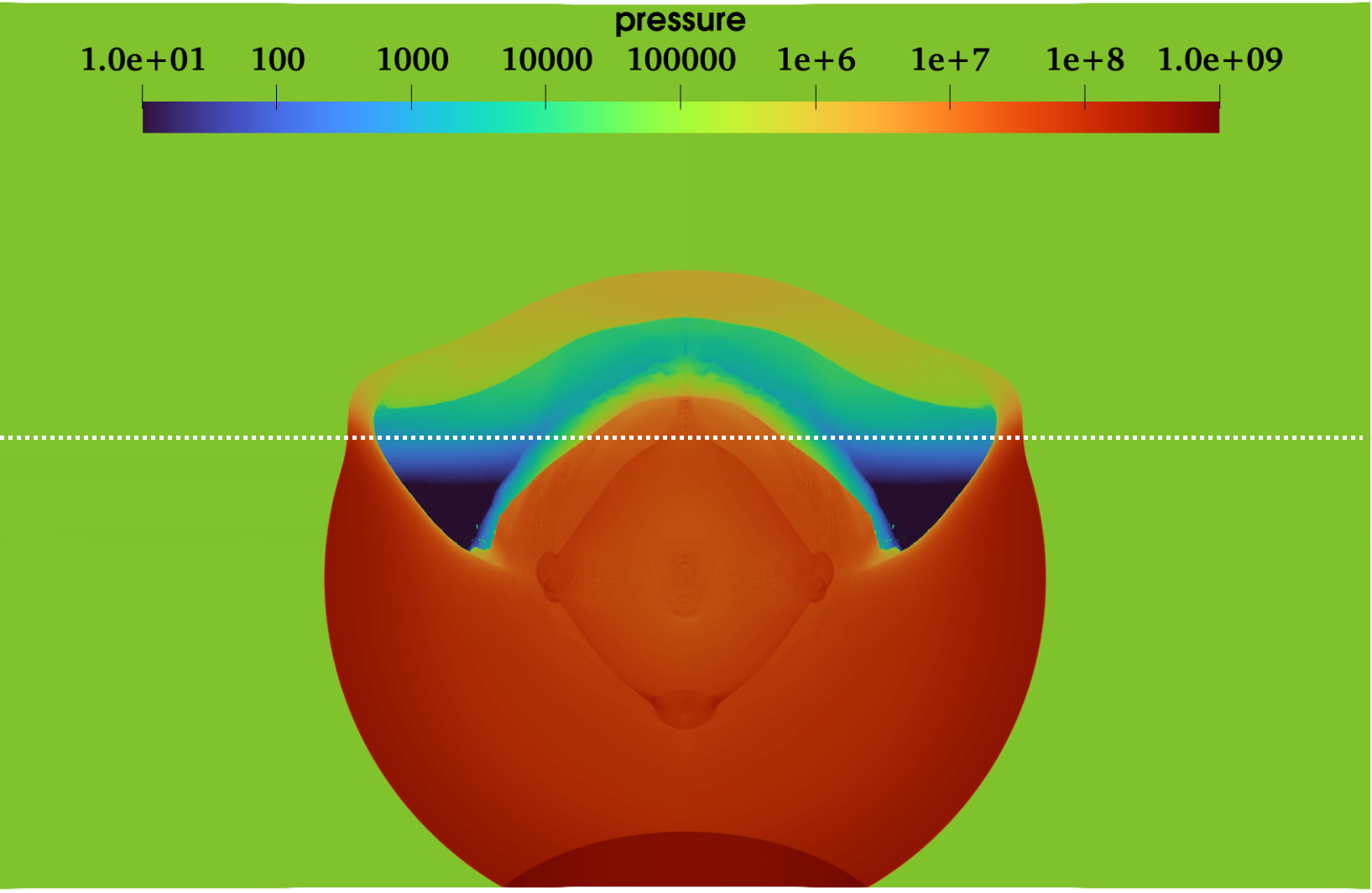}}  \\
\subfigure[pressure contours at $t=445 \mu$s]
{\includegraphics[width=0.42\textwidth]{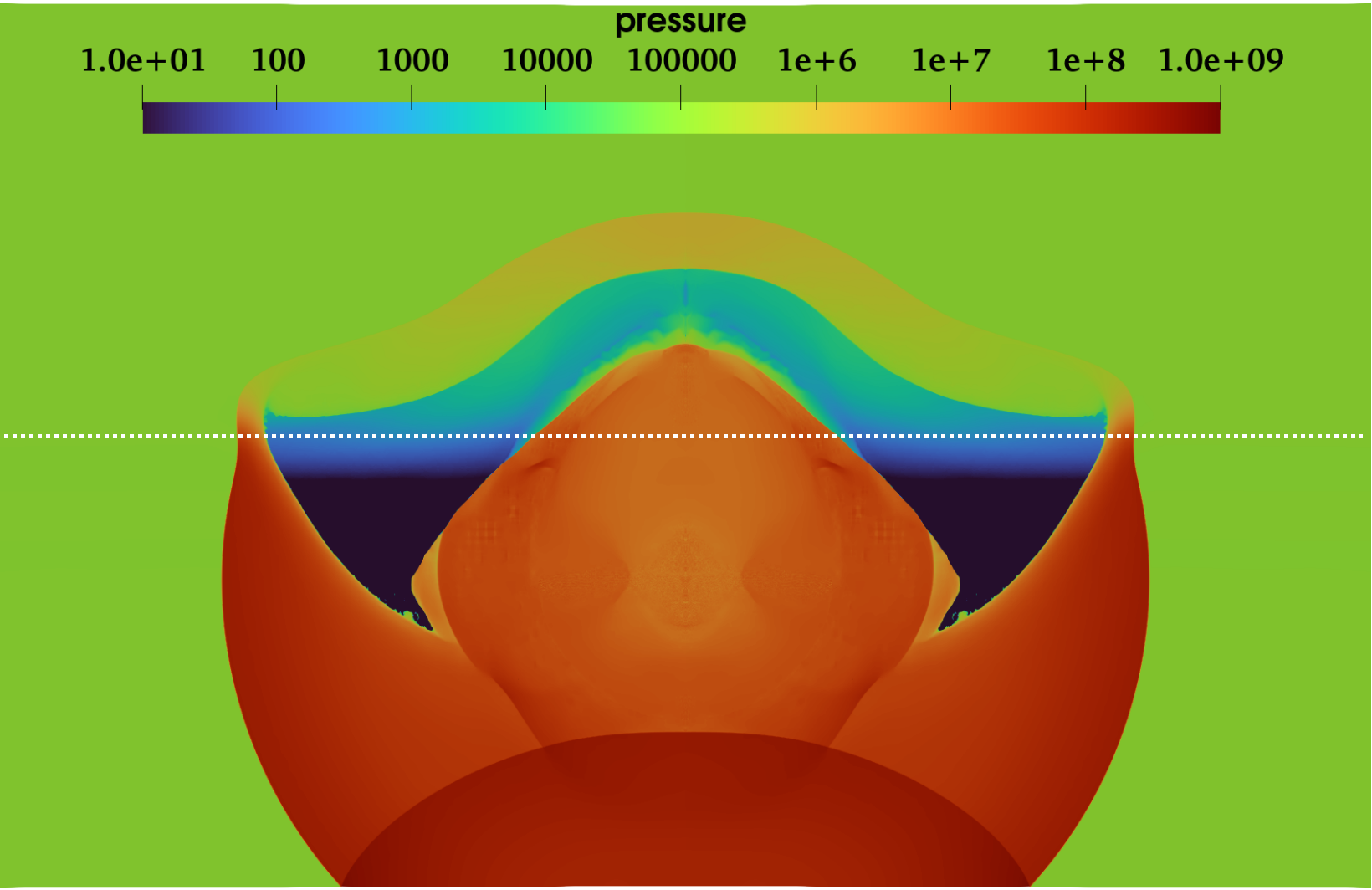}} ~~
\subfigure[pressure contours at $t=520 \mu$s]
{\includegraphics[width=0.42\textwidth]{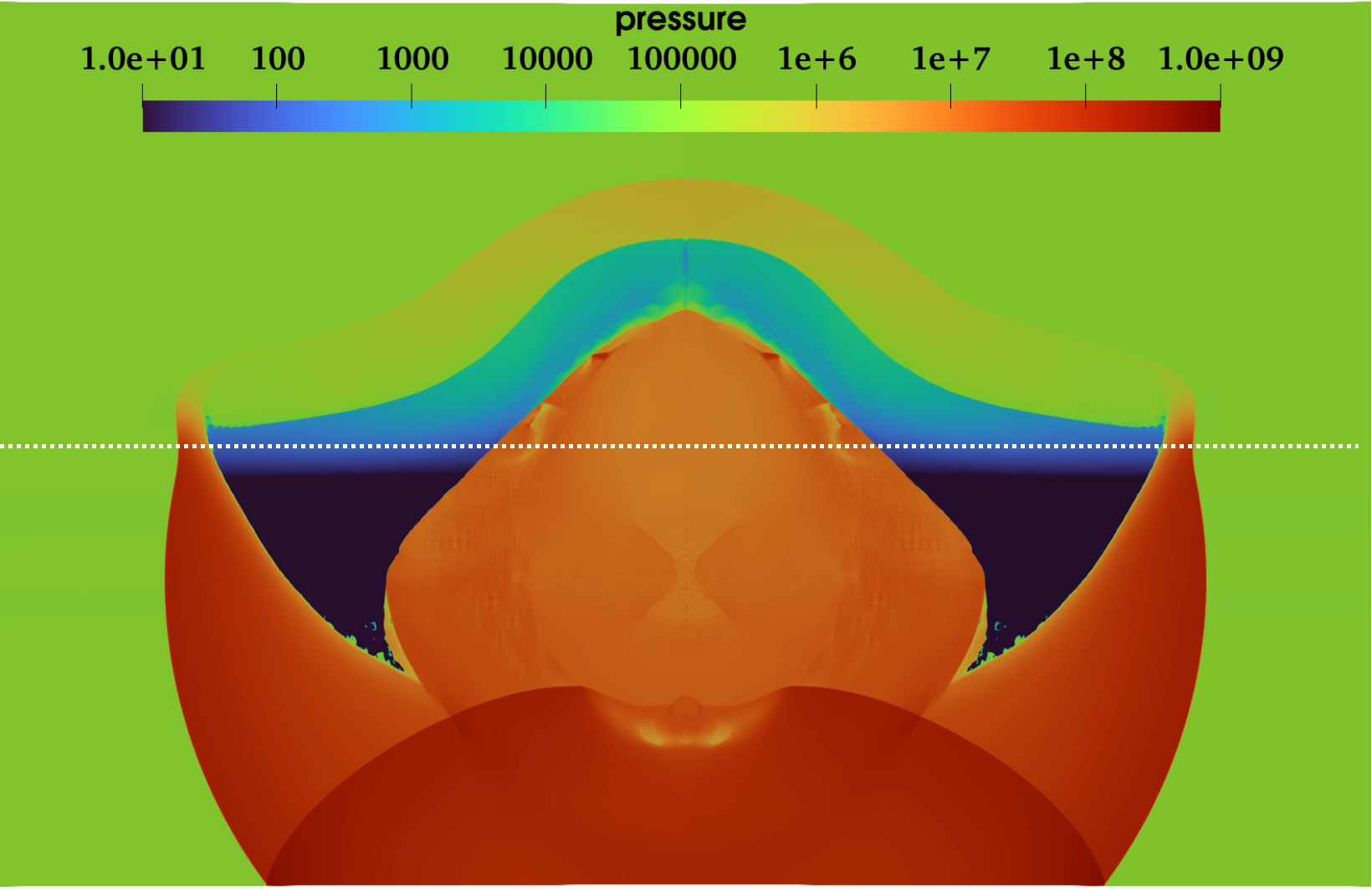}}
\caption{Shock wave contours of TNT explosion near gas-water surface.}
\label{rm:underwater2}
\end{figure}

Similar to Sec. \ref{subsec:near2d}, we adopt the diffuse interface method to capture the interfaces between the TNT, water, and air, while the new cavitation model is adopted to simulate the cavitating behavior of water when the pressure drops below the critical value $p_{sat}$. The top, left, and right boundaries are set as non-reflecting boundaries, while the bottom boundary is set as a rigid wall. The sequential snapshots of the pressure contours are depicted in Fig. \ref{rm:underwater2}, which show the whole procedures of underwater explosion in compression, expansion, and cavitation phases. The phenomena of shock wave propagation are similar to Sec. \ref{subsec:near2d} before it reaches the gas-water interface. When the shock wave reaches the gas-water interface, a transmissive wave and a reflective wave are produced. Due to the interaction of reflected compressive wave and expansive gas bubble, the pressure adjacent to the bubble will drop quickly to the critical pressure, and cavitation occurs at about 100 $\mu$s. Then the cavitation evolves below the free surface, and a very low-pressure region is formed. At last, when the shock wave arrives at the bottom rigid wall, intense reflection occurs in the adjacent area of the bottom. Similar cavitation is formed when the reflected shock wave meets the gas bubble, which leads to a very complex cavitation distribution in the whole computational region.

\begin{figure}
    \centering
    \includegraphics[width=0.6\textwidth]{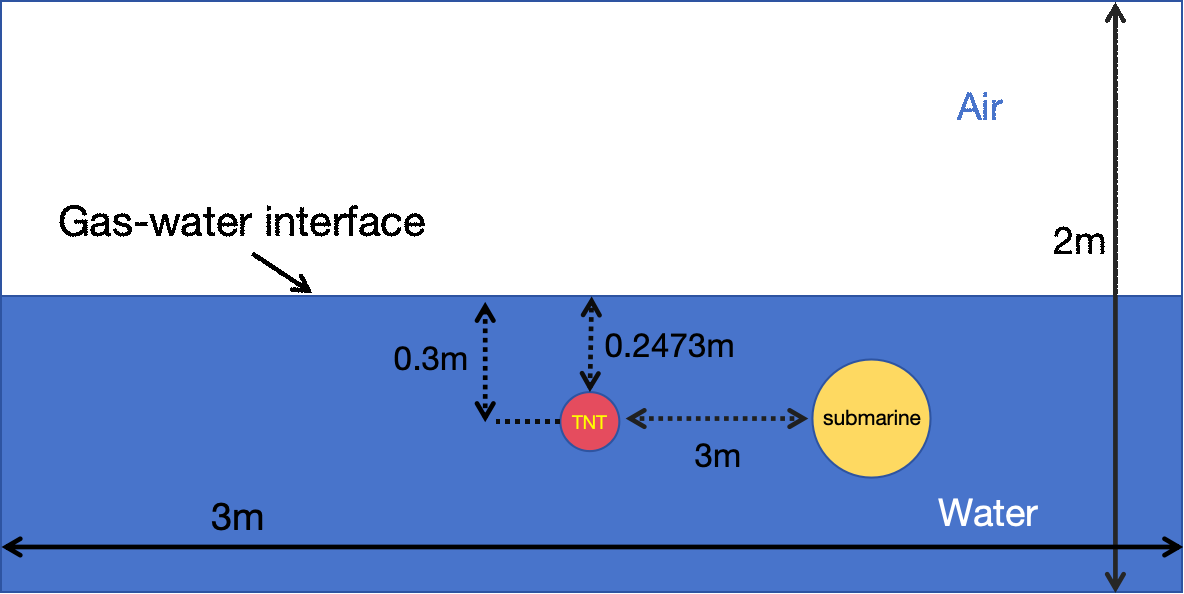}
    \caption{The Schematic diagram along the $x = 0$ to represent the initial condition of the TNT explosives(red) and the submarine(yellow).}
    \label{num:underwater3d:computation:region}
\end{figure}

\subsection{Three-dimensional underwater explosion in engineering application}\label{rm:underwater3dimensional}
In this part, we present a three-dimensional problem in engineering applications carried out on tetrahedron unstructured meshes for each phase of fluids. Similar to two-dimensional calculations, parallel computing based on the classical domain decomposition methods and the $h$-adaptive mesh method are implemented to improve the efficiency of the simulation. As shown in Fig. \ref{num:underwater3d:computation:region}, we now simulate a spherical underwater explosion near the free surface, similar to Sec. \ref{rm:underwater2dimensioanl}, where a spherical TNT explosion product is located below the gas-water interface with a radius of $r=0.0527$m. There is a submarine model located next to the TNT explosion product. We are concerned about the pressure distribution and the formation of cavitation areas near the submarine. The whole computational domain is $[0,3]\times[0,3]\times[-1,1]$ in meters. The depth of the explosion point is set as $x=0,~y=0,z=-0.3$m. We take the JWL EOS, polynomial EOS, and ideal gas EOS to simulate the TNT, water, and air, respectively. The initial conditions are
\[
 [\rho, u, v, w, p]^\top = \left\{
    \begin{array}{ll}
      [1630, ~0, ~0, ~9.5\times 10^{9}]^\top, & x^2+y^2+(z+0.3)^2 \le 0.0527, \\ [2mm]
      [1000, ~0, ~0, ~1.0\times 10^{5}]^\top, & x^2+y^2+(z+0.3)^2 > 0.0527 \text{ and } z \le 0, \\ [2mm]
      [1.0, ~0, ~0, ~1.0\times 10^{5}]^\top,  & z > 0.
    \end{array}
  \right.
\] 

\begin{figure}[htbp]
    \centering
    \subfigure[t=0.0004s]{\includegraphics[width=\textwidth]{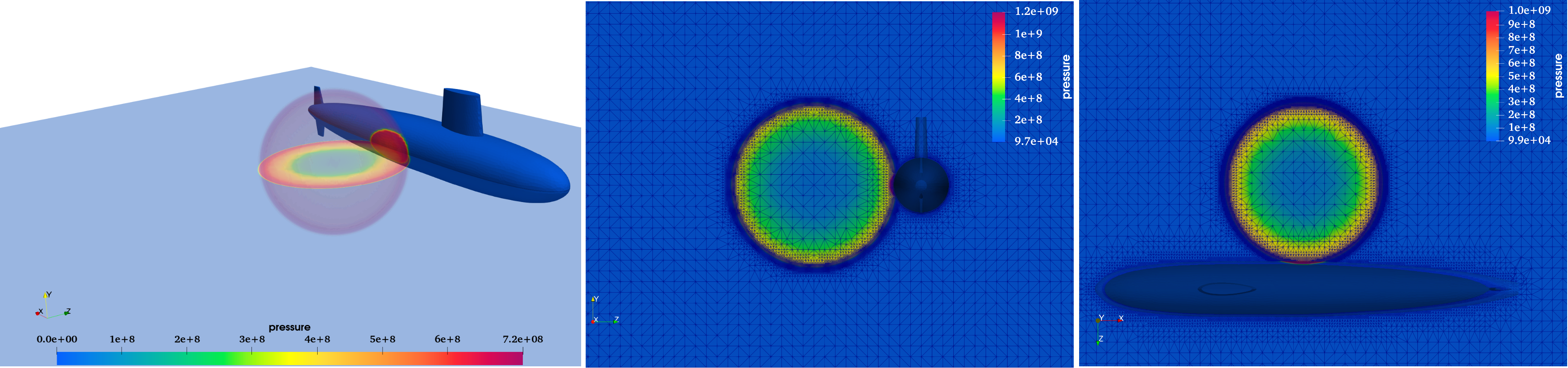}}
    \subfigure[t=0.0006s]{\includegraphics[width=\textwidth]{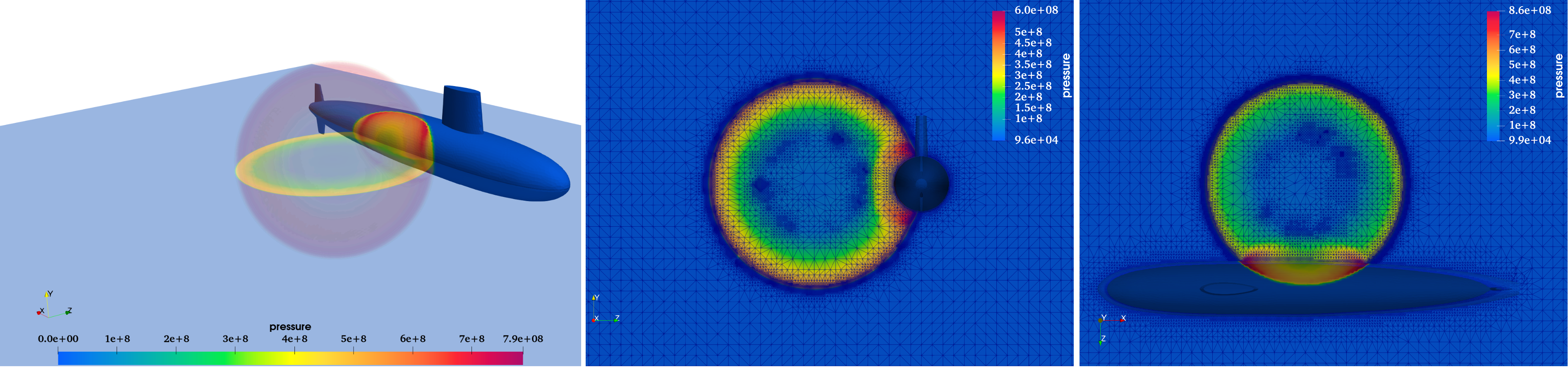}}
    \subfigure[t=0.0008s]{\includegraphics[width=\textwidth]{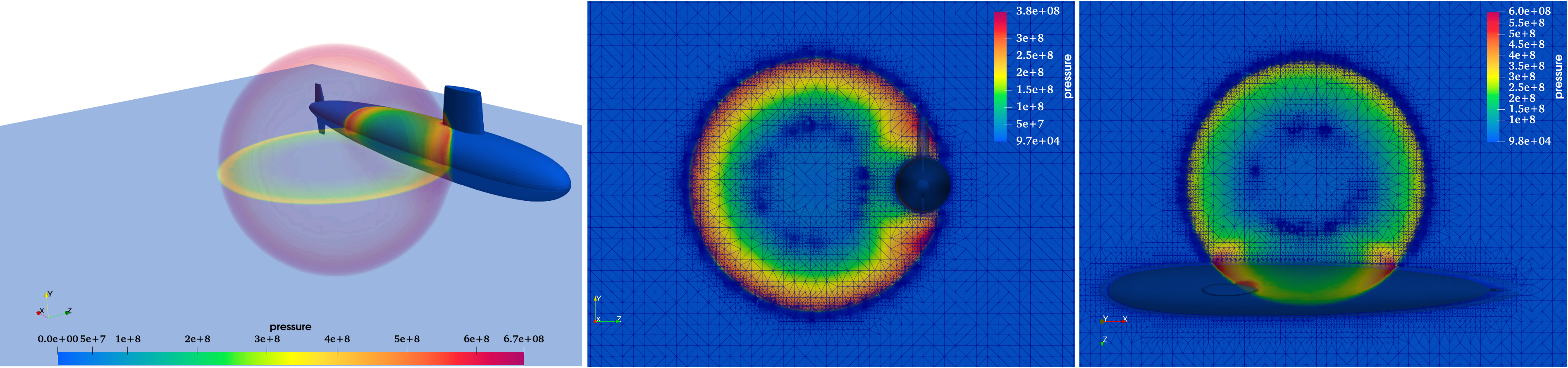}}
    \subfigure[t=0.001s]{\includegraphics[width=\textwidth]{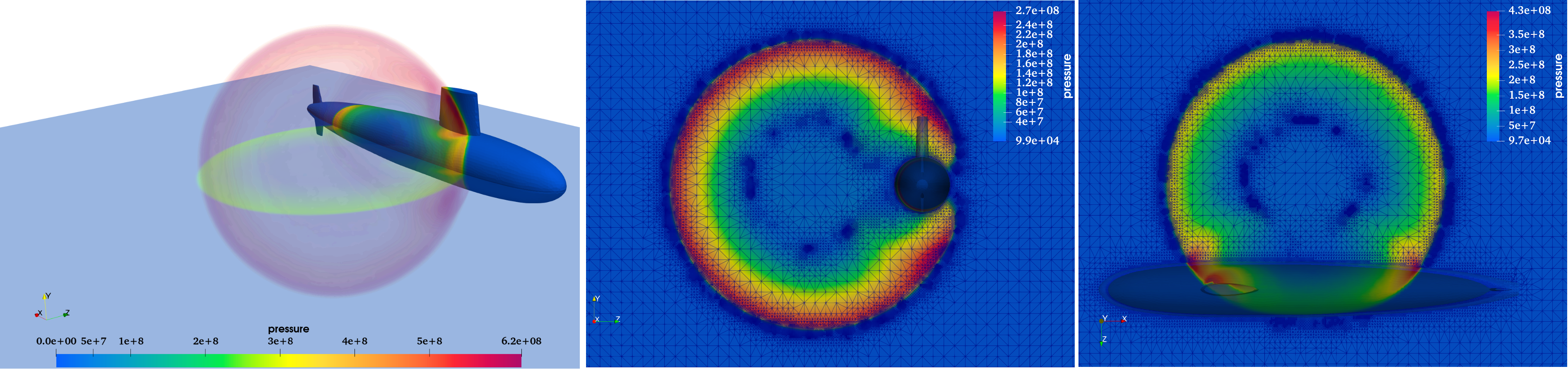}}
    \subfigure[t=0.0012s]{\includegraphics[width=\textwidth]{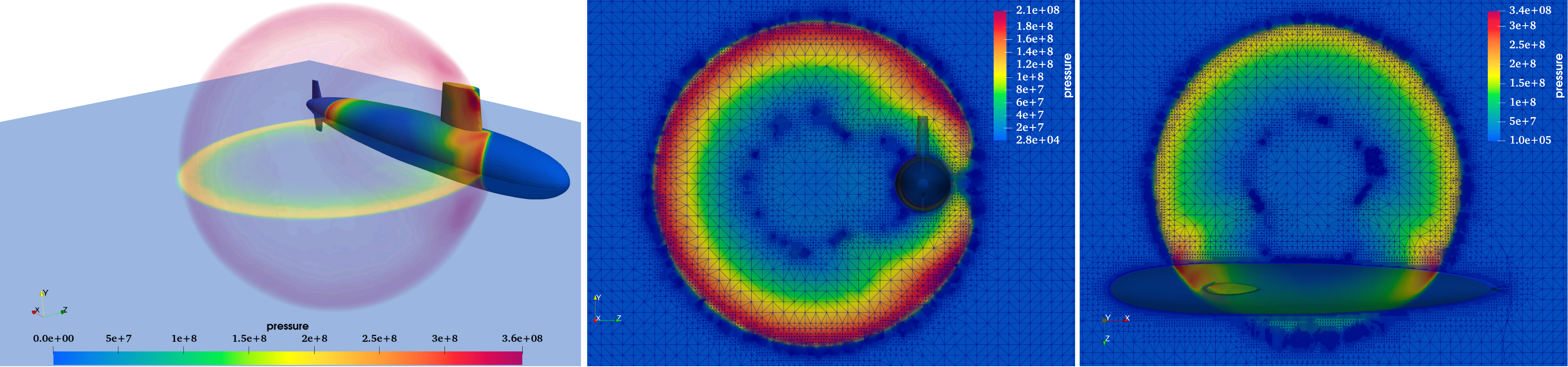}}
    \caption{Numerical results of the three-dimensional underwater explosion in example \ref{rm:underwater3dimensional}. Left column: load distribution of shock waves generated by explosives on submarines; Middle column: the cross-sectional view of pressure along the OYZ plane; Right column: the cross-section view of pressure along the OXZ plane.}
    \label{numerical:3D:1}
\end{figure}



\section{Conclusions}\label{sec:conclusion}
In this paper, we present a novel physics-informed data-driven cavitation model based on a specific Mie-Grüneisen EOS, the polynomial EOS. The neural-network-based EOS is obtained by solving an optimization problem, which combines physics loss from ODE and supervisor loss from experimental data. Our model is verified with challenging benchmark tests on the Euler equation of the single phase, comparing results with other classical cavitation models in one-dimensional problems. Various two-dimensional and three-dimensional numerical examples and their engineering applications are presented to validate their effectiveness and robustness. In multi-dimensional numerical examples, parallel computing based on classical domain decomposition methods, h-adaptive mesh refinement, and load balance are implemented to improve simulation efficiency. In our future work, we will apply our framework to other types of Mie-Grüneisen EOS and take thermodynamic behavior into account. 
\section{Acknowledgments}
\textcolor{red}{This work is financially supported by the Strategic Priority Research Program of Chinese Academy of Sciences(Grant No. XDA25010405). It is also partially supported by the National Key R$\&$D Program of China, Project Number 2020YFA0712000, the National Natural Science Foundation of China (Grant No. DMS-11771290) and the Science Challenge Project of China (Grant No.TZ2016002).}

Minsheng Huang would give special thanks to Prof. Zheng Ma at Shanghai Jiao Tong University for inspirational discussions.


\bibliographystyle{unsrt}
\bibliography{reference}
\end{document}